# Understanding the Multi-Scale and Multi-fractal Dynamics of Space Plasmas Through Tsallis Non-Extensive Statistical Theory.

G.P. Pavlos

*Department of Electrical and Computer Engineering, Democritus University of Thrace, 67100 Xanthi, Greece*
*Email: gpavlos@ee.duth.gr*

**Abstract**

*In this study it is shown that the Tsallis q-extended statistical theory was found efficient to describe faithfully the space plasmas statistics in every case, from the planetary magnetospheres, to solar corona and solar dynamics, as well as cosmic rays and cosmic stars. Moreover, new theoretical concepts and experimental results are presented concerning the space plasma complex dynamics. The significant message of theoretical and experimental issues presented here is the necessity of generalized statistical and dynamical theory for understanding the non-equilibrium dynamics and the complex character of space plasmas. The q-extension of statistics coupled to the fractal extension of dynamics are the novel and appropriate theoretical framework for the description of space plasma complexity.*

## 1. Introduction

Tsallis non-extensive statistical [1] theory can be used for a comprehensive description of space plasma dynamics, as recently we became aware of the drastic change of fundamental physical theory concerning physical systems far from equilibrium.
The dynamics of space plasma is one of the most interesting and persisting modern physical problems including the hierarchy of complex and self-organized phenomena such as: intermittent turbulence, fractal structures, long range correlations, far from equilibrium phase transitions, anomalous diffusion – dissipation and strange kinetics, reduction of dimensionality etc. [2-9]. These plasma phenomena are related with the oldest and most dominant problems of space or laboratory plasmas related is the magnetic reconnection process.
In this direction Chang [3] introduced the generalized Wilsonian approach [10] for the study of space plasma. According to this point of view, plasma dynamics must be understood as a far from equilibrium multi-scale turbulent and multi-fractal intermittent process described by the non-equilibrium renormalization group theory (RGT). Recently, similar theoretical concepts were supported also by Consolini et al [7], Lui[11], Sharma et al [6], Sitnov et al [12], Zelenyi et al [2], Milovanov et al [4], Abramenco [8], Pavlos et al [9] and others. The holistic and self-organizing character of space plasma processes was supported by Pavlos et al [9] after a deep and impressive influence by the scientific work of Prigogine [13], Nicolis [14] and Haken [15]. More than other scientists, Prigogine, as he was deeply inspired by the arrow of time and the chemical complexity, supported the marginal point of view that the dynamical determinism of physical reality is produced by an underlying ordering process of entirely holistic and probabilistic character at every physical level. If we accept this extreme scientific concept, then we must accept also for space plasma the new point of view, that the MHD or kinetic theory of plasma are inefficient to describe sufficiently the emerging complex character. However resent evolution of the physical theory centered on non-linearity and fractality shows that the Prigogine point of view was so that much extreme as it was considered at the beginning.

After all, Tsallis $q$ – extension of statistics [1] and the fractal extension for dynamics of complex systems as it has been developed by Notalle [16], El Naschie [17], Castro [18], Tarasov [19], Zaslavsky [20], Milovanov [4], El Nabulsi [21], Cresson [22], Coldfain [23], Chen [24], and others scientists, they are the double face of a unified novel theoretical framework, and constitute the appropriate base for the modern study of space plasmas. Also the $q$ – statistics is related to the underlying fractal dynamics of the non-equilibrium states. The conceptual novelty of complexity theory embraces all of the physical reality from equilibrium to non-equilibrium states. This is noticed by Castro [18] as follows: *"…it is reasonable to suggest that there must be a deeper organizing principle, from small to large scales, operating in nature which might be based in the theories of complexity, non-linear dynamics and information theory which dimensions, energy and information are intricately connected." [18]*.

This novel point of view is in agreement with modern transformation of scientific concepts especially for space plasma processes, where magnetic reconnection and magnetic energy dissipation is one of the main great problems of space plasma physics [25, 26]. The physical explanation of MHD reconnection by the kinetic theory of the magnetized and collisionless plasma was steadily attempted the last decades [27, 28]. New blood was infused by the Chang theory [3] of non-equilibrium critical dynamics. The concept of low dimensional chaos and strange attractor for the interpretation of the solar plasma dynamics was introduced by Weiss [29] and Ruzmakin [5], by Baker [30] and Vassiliadis [31] for the magnetospheric space plasmas dynamics and by Pavlos [9] for the solar and magnetospheric space plasmas. Tsallis non-extensive statistical theory, introduced as a multifractal extension of Boltzmann – Gibbs statistics [1], was used for the description of the space plasma dynamics by Burlaga [32], Ferri et al [33], Freitas and De Medeiros [34], Pavlos et al [9] and Karakatsanis et al [2010, 2012]. In this study Tsallis theory is used as the new theoretical framework for the physical interpretation and comprehension of the non-equilibrium space plasmas dynamics.

## 2. Theoretical Concepts

Space plasmas are far from equilibrium distributed dynamical systems including as subsystems: solar convection zone and solar corona, solar wind and planetary magnetospheres, interstellar medium etc. Globally and locally, the space plasma system is a highly organized and magnetized system with strong temperature density fields and other transport magnitude gradients, sustained by the inner sun activity and its dynamo dynamic in the convection zone. In every case, the space plasma system includes rich internal non-linear dynamics coupled with its environment through matter – energy input – output process. In every case, the input rates of matter – energy from the local space plasma environment constitute the control parameters of the internal dynamics.

### 2.1 The new concepts of complexity theory

For complex systems near equilibrium the underlying dynamics and the statistics are Gaussian as it is caused by a normal Langevin type stochastic process with a white noise Gaussian component. The normal Langevin stochastic equation corresponds to the probabilistic description of dynamics by the well-known normal Fokker – Planck equation. For Gaussian processes only the moments-cumulants of first and second order are non-zero, while the central limit theorem inhibits the development of long range correlations and macroscopic self-organization, as any kind of fluctuation quenches out exponentially to the normal distribution. Also at equilibrium, the dynamical attractive phase space is practically infinite dimensional as the system state evolves in all dimensions according to the famous ergodic theorem of Boltzmann – Gibbs statistics. However, in Tsallis $q$ – statistics even for the case of $q = 1$ (corresponding to Gaussian process) the non-extensive character permits the development of long range correlations produced by equilibrium phase transition multi-scale processes according to the Wilson RGT [10]. From this point of view, the classical mechanics (particles and fields), including also general relativity theory, as well as the quantum mechanics – quantum field theories, all of them are nothing else than a near thermodynamical equilibrium approximation of a wider theory of physical reality, characterized as complexity theory. This theory can be related with a globally acting ordering process which produces the $q$ – statistics and the fractal extension of dynamics classical or quantum.

Generally, the experimental observation of a complex system presupposes non-equilibrium process of the physical system which is subjected to observation, even if the system lives thermodynamically near to equilibrium states. Also experimental observation includes discovery and ascertainment of correlations in space and time, as the spatio-temporal correlations are related or they are caused by from the statistical mean values fluctuations. The theoretical interpretation prediction of observations as spatial and temporal correlations – fluctuations is based on statistical theory which relates the microscopic underling dynamics with the macroscopic observations indentified to statistical moments and cumulants. Moreover, it is known that statistical moments and cumulants are related to the underlying dynamics by the derivatives of the partition function ($Z$) to the external source variables ($J$) [37].

From this point of view, the main problem of complexity theory is how to extend the knowledge from thermodynamical equilibrium states to the far from equilibrium physical states. The non-extensive $q$ – statistics introduced by Tsallis [1] as the extension of Boltzmann – Gibbs equilibrium statistical theory is the appropriate base for the non-equilibrium extension of complexity theory. The far from equilibrium $q$ – statistics can produce the $q$-partition function ($Z_q$) and the corresponding $q$ – moments and cumulants, in correspondence with Boltzmann – Gibbs statistical interpretation of thermodynamics.

The miraculous consistency of physical processes at all levels of physical reality, from the macroscopic to the microscopic level, as well as the inefficiency of existing theories to produce or to predict the harmony and hierarchy of structures inside structures from the macroscopic or the microscopic level of cosmos. This completely supports or justifies such new concepts as that indicated by Castro [18]: *"of a global ordering principle or that indicated by Prigogine, about the becoming before being at every level of physical reality."* The problem however with such beautiful concepts is how to transform them into an experimentally testified scientific theory.

For space plasmas out of equilibrium the development of long range correlations in various complex structures of multi-scale and multi-fractal character, the corresponding central problem concerns the inefficiency of the classic theory, which is based on local particle- field interactions, to confront the experimentally verified space plasma complexity. Although we haven't unobjectionable or undeniable proof and certificate, we believe that classical plasma theory cannot solve this problem. In reality the theoretical problem of space plasmas complex dynamics is included in the wider problem of random fields and their novel non-equilibrium statistical theory [3, 7, 38].

The Feynman path integral formulation of quantum theory after the introduction of imaginary time transformation by the Wick rotation indicates the inner relation of quantum dynamics and statistical mechanics [39, 62]. In this direction it was developed the stochastic and chaotic quantization theory [40-43], which opened the road for the introduction of the macroscopic complexity and self-organization in the region of fundamental quantum field physical theory. The unified character of macroscopic and microscopic complexity is moreover verified by the fact that the $n$ – point Green functions produced by the generating functional $W(J)$ of QFT after the Wick rotation can be transformed to $n$ – point correlation functions produced by the partition function $Z(J)$ of the statistical theory. This indicates in reality the self-organization process underlying the creation and interaction of elementary particles, similarly to the development of correlations in complex systems and classical random fields [40]. For this reason lattice theory describes simultaneously microscopic and macroscopic complexity [10, 39].

In this way, instead of explaining the macroscopic complexity by a fundamental physical theory such as QFT, Superstring theory, M-theory or any other kind of fundamental theory we become witnesses of the opposite fact, according to what Prigogine was imagining. That is, macroscopic self-organization process and macroscopic complexity install their kingdom in the heart of reductionism and fundamentalism of physical theory. The Renormalizable field theories with the strong vehicle of Feynman diagrams that were used for the description of high energy interactions or the statistical theory of critical phenomena and the nonlinear dynamics of plasmas [44] lose their efficiency when the complexity of the process scales up [10].

Many scientist as Chang [3], Zelenyi [2], Milovanov [4], Ruzmaikin [5], Abramenko [8], Lui [11], Pavlos [9], in their studies indicate the statistical non-extensivity as well as the multi-scale, multi-fractal and anomalous – intermittent character of fields and particles in the space plasma dynamics. These results verify the concept that space plasma dynamics are part of the more general theory of fractal dynamics which has been developed rapidly the last years. Fractal dynamics are the modern fractal extension of physical theory in every level. On the other side the fractional generalization of modern physical theory is based on fractional calculus: fractional derivatives or integrals or fractional calculus of scalar or vector fields and fractional functional calculus [19, 24]. It is very impressive the efficiency of fractional calculus to describe complex and far from equilibrium systems which display scale-invariant properties, turbulent dissipation and long range correlations with memory preservation, while these characteristics cannot be illustrated by using traditional analytic and differentiable functions, as well as, ordinary differential operators. Fractional calculus permits the fractal generalization of Lagrange – Hamilton theory of Maxwell equations and Magnetohydrodynamics, the Fokker – Planck equation Liouville theory and BBGKI hierarchy, or the fractal generalization of QFT and path integration theory [19, 21-24].

According to the fractal generalization of dynamics and statistics we conserve the continuity of functions but abolish their differentiable character based on the fractal calculus which are the non-differentiable generalization of differentiable calculus. At the same time the deeper physical meaning of fractal calculus is the unification of microscopic and macroscopic dynamical theory at the base of the space – time fractality [16, 17, 24, 45, 46, 47]. Also ,the space-time is related to the fractality – multi-fractality of the dynamical phase – space, whish can be manifested as non-equilibrium complexity and self-organization.

Moreover fractal dynamics leads to a global generalization of physical theory as it can be related with the infinite dimension Cantor space, as the microscopic essence of physical space – time, the non-commutative geometry and non-commutative Clifford manifolds and Clifford algebra, or the p-adic physics [17, 18, 21, 48, 49]. According to these new concepts introduced the last two decades at every level of physical reality we can describe in physics complex structure which cannot be reduced to underlying simple fundamental entities or underlying simple fundamental laws. Also, the non-commutative character of physical theory and geometry indicates [49, 50] that the scientific observation is nothing more than the observation of undivided complex structures in every level. Cantor was the founder of the fractal physics creating fractal sets by contraction of the homogenous real number set, while on the other side the set of real numbers can be understood as the result of the observational coarse graining [48, 51, 52]. From a philosophical point of view the mathematical forms are nothing else than self-organized complex structures of the mind-brain, in self-consistency with all the physical reality. On the other side, the generalization of Relativity theory to scale relativity by Nottale [16] or Castro [18] indicates the unification of microscopic and macroscopic dynamics through the fractal generalization of dynamics.

After all, we conjecture that the macroscopic self-organization related with the novel theory of complex dynamics, as they can be observed at far from equilibrium dynamical physical states, are the macroscopic emergence result of the microscopic complexity which can be enlarged as the system arrives at bifurcation or far from equilibrium critical points. That is, far from equilibrium the observed physical self-organization manifests the globally active ordering principle to be in priority from local interactions processes. We could conjecture that is not far from thruth the concept that local interactions themselves are nothing else than local manifestation of the holistically active ordering principle. That is what until now is known as fundamental lows is the equilibrium manifestation or approximation of the new and globally active ordering principle. This concept can be related with the fractal generalization of dynamics which is indentified with the dynamics of correlations supported by Prigogine [13], Nicolis [14] and Balescu [53], as the generalization of Newtonian theory. This conjecture concerning the fractal unification of macroscopic and microscopic dynamics at can be strongly supported by the Tsallis nonextensive q-statistics theory which is verified almost everywhere from the microscopic to the macroscopic level [1, 18]. From this point of view it is reasonable to support that the q-statistics and the fractal generalization of space plasma dynamics is the appropriate framework for the description of their non-equilibrium complexity.

## 2.2 The conspiracy of mathematics and physical reality.

The microscopic description, the magnetohydrodynamical (MHD) or the kinetic approximation of space plasma systems corresponds to non-linear partial differential equations of the general type:

$$\frac{\partial \vec{U}(\vec{x},t)}{\partial t} = \vec{F}(\vec{u}, \vec{\lambda}) \tag{1}$$

where $u$ belongs to a infinite dimensional state (phase) space which is a Hilbert functional space. Among the various control parameters, the plasma Reynold number is the one which controls the quiet static or the turbulent plasma states. Generally the control parameters measure the distance from the thermodynamical equilibrium as well as the critical or bifurcation points of the system for given and fixed values, depending upon the global mathematical structure of the dynamics. As the system passes its bifurcation points a rich variety of spatio-temporal patterns with distinct topological and dynamical profiles can be emerged such as: limit cycles or torus, chaotic or strange attractors, turbulence, Vortices, percolation states and other kinds of complex spatiotemporal structures [3, 32, 33, 47, 54, 56, 57, 63, 70, 71, 73].

For space plasmas the mathematical systems equation [1] corresponds to different levels of plasma description as magnetohydromagnetic equations (MHD) or kinetics equations. MHD equations are the extension of Navier – Stoke equations for electrically conducting continuous media including electric ($\vec{E}$) and magnetic ($\vec{B}$) fields and electric currents ($\vec{J}$) as follows:

$$\rho \frac{d\vec{V}}{dt} = -\vec{\nabla}p + \vec{J}x\vec{B} + vp\nabla^2\vec{v} \tag{2}$$

where $p$ is the pressure, $\rho$ is the density mass and $\nu$ is the kinematic viscosity. The hydrodynamic component of space plasma is connected with its electromagnetic components by the Maxwell's equations:

$$\vec{\nabla} X \vec{B} = \mu_0 \vec{J}, \quad \vec{\nabla} \cdot \vec{B} = 0, \quad \partial \vec{B}/\partial t = -\vec{\nabla} X \vec{E} \quad (3)$$

and the Ohm's law:

$$\vec{E} + \vec{V} X \vec{B} = n\vec{J} \quad (4)$$

where $n$ is the electric resistivity.

A system of coupled equations which describe the evolution of $\vec{V}$ and $\vec{B}$ may be obtained by using equations (2-4) by eliminating the electric field intensity $\vec{E}$ as follows:

$$\frac{\partial \vec{V}}{\partial t} = -\frac{1}{\rho}\vec{\nabla}\rho + \frac{1}{\rho}\vec{J}X\vec{B} - \nu\vec{\nabla}X\vec{\Omega} + \vec{V}X\vec{\Omega} - \vec{\nabla}\frac{V^2}{2} \quad (5)$$

$$\frac{\partial \vec{B}}{\partial t} = \vec{\nabla}X(\vec{V}X\vec{B}) - n\vec{\nabla}X\vec{J} \quad (6)$$

where $\vec{\Omega} = \vec{\nabla}X\vec{V}$ is the vorticity vector of the magnetized fluid.

The last system of MHD equations constitute a nonlinear system of the general type (1) as the $\vec{U}(\vec{x},t)$ field corresponds to the $\vec{V}(\vec{x},t)$ and $\vec{B}(\vec{x},t)$ fields. The non-linearity of MHD equations is related to complex dynamics including the plasma turbulence and other multi-scale complex and self-organization phenomena, related to the scale invariant character of MHD equations. Scale invariance means that the MHD equations are invariant under the scaling transformations:

$$\left.\begin{array}{ll} \vec{X}' = \lambda\vec{X}, & \vec{V}' = \lambda^{a/3}\vec{V} \\ t' = \lambda^{1-a/3}t, & \vec{B}' = \lambda^{a/3}\vec{B} \\ (p/\rho)' = \lambda^{2a/3}(p/\rho) & \end{array}\right\} \quad (7)$$

[55, 56]. The scale invariance of MHD systems couples the plasma theory with the renormalization group theory (RG) [10]. The RG theory is the road to the hidden complexity of the plasma dynamics.

That is at bifurcation points the non-linear mathematical system reveals spontaneous formation of spatio-temporal structures and sudden qualitative changes of the plasma state. According to Kovacs [57], in the vicinity of a bifurcation point the non-linear system becomes extremely sensitive to small noise components which can lead to symmetry breaking and pattern formation such as: coherent patterns with long-range spatial order, long-range temporal order (chaotic synchronization), early (or few-mode) turbulence, fully developed turbulence or defect turbulence and detect chaos as well as spatiotemporal intermittency [38]. As we indicate in the next sessions non-linear dynamics couples the physical process of complex systems with the topological and other geometrical properties of the physical space time reality.

Until now, the previous deterministic dynamics for the continuous plasma states is only an approximation, as the macroscopic variables correspond to mean values of a coarse – graining process. For the classical point of view the determination of plasma dynamics can be based at the microscopic level where point charged particles and microscopic fields satisfy the deterministic equations of motion:

$$\frac{d\vec{X}_i}{dt} = \vec{V}_i, \quad \frac{d\vec{V}_i}{dt} = \frac{q_i}{m_i}\left(E^M + \vec{V}_i X B^M\right) = \frac{1}{m_i}F^M \quad (8)$$

while the distribution function:

$$N_a(\vec{x},\vec{u},t) = \sum_{1 \leq i \leq N_a} \delta[\vec{x} - \vec{x}_i(t)]\delta[\vec{u} - \vec{v}i(t)] \quad (9)$$

can be used to estimate the field sources:

$$\vec{\nabla} \cdot \vec{E}^M = \left(\sum q \int N(\vec{x},\vec{v})d\vec{v}\right) / \varepsilon_0 \quad (10a)$$

$$\vec{\nabla} \cdot \vec{B}^M = \mu_0 \varepsilon_0 \partial \vec{E}^M / \partial t + \mu_0 \sum q \int \vec{u} N(\vec{x},\vec{u},t)d\vec{v} \quad (10b)$$

$$\vec{\nabla} \cdot \vec{E}^M = -\frac{\partial \vec{B}^M}{\partial t}, \quad \vec{\nabla} \cdot \vec{B}^M = 0 \quad (10c)$$

The distribution function $N(\vec{x},\vec{u},t)$ satisfies the Klimontovich – Dupree equation

$$\frac{\partial N}{\partial t} + \vec{u} \cdot \frac{\partial N}{\partial \vec{x}} + \frac{F^M}{m} \cdot \frac{\partial N}{\partial \vec{u}} = 0 \quad (11)$$

This is the generalized Liouville equation from which the BBGKY hierarchy of the stochastic description can be produced [58].

Equations [8-11] constitute another (deterministic) realization of the general non-linear mathematical system described by equation [1]. In the classical description of plasmas the stochastic character of macroscopic MHD or kinetic equations is caused by the coarse – graining process of the estimation of the first, second or higher moments and correlations functions. However, the chaotic dynamics caused by non-linearity of the microscopic equations embeds the plasma dynamics in the wider perspective of fractal dynamics [2, 4, 19, 20], as we present in the next sections, where stochasticity is rooted ontologically at the foundation of physical reality itself.

Generally chaotic solutions of the mathematical system (1) force to correspond it to a stochastic non-linear stochastic system:

$$\frac{\partial \vec{u}}{\partial t} = \vec{\Phi}(\vec{u},\vec{\lambda}) + \vec{\delta}(\vec{x},t) \quad (12)$$

where $\vec{\delta}(\vec{x},t)$ corresponds to the random force fields produced by strong chaoticity [36, 59].

The non-linear mathematical systems (1-2) include give mathematical solutions which can represent plethora of non-equilibrium physical states included in mechanical or chemical and other physical systems and of course in space plasma, which is the central theme of this study. For space plasmas the general Langevin stochastic equation corresponds to the kinetic plasma equations:

$$\frac{\partial f}{\partial t} = \vec{V} \cdot \vec{\nabla}_x f_a + \frac{q_a}{m_d}(\vec{E} + \vec{V}X\vec{B}) \cdot \vec{\nabla}_v f_a^t = \delta f \quad (13)$$

$$\frac{\partial \vec{E}}{\partial t} = (\mu_0 \vec{\nabla}X\vec{B} - \frac{1}{\rho}\sum n_a q_a \int \vec{V}f_a d\vec{V}) / \mu_0 \varepsilon_0 + \delta \vec{E} \quad (14)$$

$$\frac{\partial \vec{B}}{\partial t} = -\vec{\nabla}X\vec{E} + \delta \vec{B} \quad (15)$$

$$\nabla \cdot \vec{E} = \left(\sum_a n_a q_a \int f_a d\vec{V}\right) / \varepsilon_0 \quad (16)$$

where $f_a(\vec{x},\vec{v},t) = \langle Na \rangle$ is the one-particle distribution, while $\vec{E}$ and $\vec{B}$ are the average values ($<\vec{E}^m>,<\vec{B}^m>$) of the microscopical fields ($\vec{E}^M, \vec{B}^M$). The fluctuations ($\delta f, \delta \vec{E}, \delta \vec{B}$) correspond to the microscopic random components of the dynamical systems. Equations (2), (5), (13), can be obtained by the Liouville equations of the self-consistence plasma field and particle equations [58]. The random components ($\delta f, \delta \vec{E}, \delta \vec{B}$) correspond to the BBGKY hierarchy:

$$\frac{\partial f_q}{\partial t} = [H_q, f_a] + S_q, q = 1, 2, ..., N \qquad (17)$$

where $f_q$ is the $q$–particle distribution function, $H_q$ is the $q$–th approximation of the Hamiltonian $q$–th correlations and $S_q$ is the statistical term including correlations of higher than q-orders [44, 59].

The non-linear mathematical systems correspond to the new science known today as complexity science. This new science has a universal character, including an unsolved scientific and conceptual controversy which is continuously spreading in all directions of the physical reality and concerns the integrability or computability of the dynamics [60]. This is something supported by many scientists after the Poincare discovery of chaos and non-integrability as is it shown in physical sciences by the work of Prigogine, Nicolis, Yankov and others [13, 14, 60]. Non-linearity and chaos is the top of a hidden mountain including new physical and mathematical concepts as fractal calculus, p-adic physical theory, non-commutative geometry, fuzzy anomalous topologies fractal space-time etc [17-24, 48-50]. These new mathematical concepts obtain their physical power when the physical system lives far from equilibrium.

After this, by following the traditional point of view of physical science we arrive at the central conceptual problem of complexity science. That is, how is it possible that the local interactions in a spatially distributed physical system can cause long range correlations or how they can create complex spatiotemporal coherent patterns as the previous non-linear mathematical systems reveal, when they are solved arithmetically, or in situ observations reveal in space plasma systems. In the case of space plasma dynamics the above questions make us to ask how the sunspots, the solar flares, the magnetospheric sub-storms or the plasma turbulence and the magnetic reconnection all of them can be explained and described by local interactions of particles and fields. At a first glance the problem looks simple supposing that it can be explained by the self-consistent particle-fields classical interactions. However the existed rich phenomenology of space plasma phenomena reveals the non-classical complex character of the universal non-equilibrium critical dynamics [3,7].

For better understandings of the new concepts we follow the road of non-equilibrium statistical theory [3, 38]

The stochastic Langevin equations (11, 13, 17) can take the general form:

$$\frac{\partial u_i}{\partial t} = -\Gamma(\vec{x})\frac{\delta H}{\delta u_i(\vec{x}, t)} + \Gamma(\vec{x}) n_i(\vec{x}, t) \qquad (18)$$

where $H$ is the Hamiltonian of the system, $\delta H / \delta u_i$ its functional derivative, $\Gamma$ is a transport coefficient and $n_i$ are the components of a Gaussian white noise:

$$\left.\begin{array}{l} < n_i(\vec{x}, t) > = 0 \\ < n_i(\vec{x}, t) n_j(\vec{x}', t') > = 2\Gamma(\vec{x})\delta_{ij}\delta(\vec{x} - \vec{x}')\delta(t - t') \end{array}\right\} \qquad (19)$$

[3, 38, 59, 61]. The above stochastic Langevin Hamiltonian equation (18) can be related to a probabilistic Fokker – Planck equation [3]:

$$\frac{1}{\Gamma(\vec{x})}\frac{\partial P}{\partial t} = \frac{\delta}{\delta \vec{u}} \cdot \left(\frac{\delta H}{\delta \vec{u}} P + \frac{\delta}{\delta \vec{u}}[\Gamma(\vec{x}) P]\right) \qquad (20)$$

where $P = P(\{u_i(\vec{x},t)\},t)$ is the probability distribution function of the dynamical configuration $\{u_i(\vec{x},t)\}$ of the system at time $t$. The solution of the Fokker – Planck equation can be obtained as a functional path integral in the state space $\{u_i(\vec{x})\}$:

$$P(\{u_i(\vec{x})\},t) \simeq \int \Delta \vec{Q} \exp(-S) P_0(\{u_i(\vec{x})\},t_0) \qquad (21)$$

where $P_0(\{u_i(\vec{x})\},t_0)$ is the initial probability distribution function in the extended configuration state space and $S = i\int L dt$ is the stochastic action of the system obtained by the time integration of it's stochastic Lagrangian (L) [3, 15]. The stationary solution of the Fokker – Planck equation corresponds to the statistical minimum of the action and corresponds to a Gaussian state:

$$P(\{u_i\}) \sim \exp\left[-(1/\Gamma) H(\{u_i\})\right] \qquad (22)$$

The path integration in the configuration field state space corresponds to the integration of the path probability for all the possible paths which start at the configuration state $\vec{u}(\vec{x},t_0)$ of the system and arrive at the final configuration state $\vec{u}(\vec{x},t)$. Langevin and F-P equations of classical statistics include a hidden relation with Feynman path integral formulation of QM [3, 39, 40, 62]. The F-P equation can be transformed to a Schrödinger equation:

$$i\frac{d}{dt}\hat{U}(t,t_0) = \hat{H} \cdot \hat{U}(t,t_0) \qquad (23)$$

by an appropriate operator Hamiltonian extension $H(u(\vec{x},t)) \Rightarrow \hat{H}(\hat{u}(\vec{x},t))$ of the classical function $(H)$ where now the field $(u)$ is an operator distribution [3,61]. From this point of view, the classical stochasticity of the macroscopic Langevin process can be considered as caused by a macroscopic quandicity revealed by the complex system as the F-K probability distribution $P$ satisfies the quantum relation:

$$P(u,t | u,t_0) = \langle u | \hat{U}(t,t_0) | u_0 \rangle \qquad (24)$$

This generalization of classical stochastic process as a quantum process could explain the spontaneous development of long-range correlations at the macroscopic level as an enlargement of the quantum entanglement character at critical states of complex systems. This interpretation is in faithful agreement with the introduction of complexity in sub-quantum processes and the chaotic – stochastic quantization of field theory [40-43], as well as with scale relativity principles [16, 18, 47] and fractal extension of dynamics [17, 19, 21-24] or the older Prigogine self-organization theory [13]. Here, we can argue in addition to previous description that quantum mechanics is subject gradually to a fractal generalization [18, 19, 21-23]. The fractal generalization of QM-QFT drifts along also the tools of quantum theory into the correspondent generalization of RG theory or path integration and Feynman diagrams. This generalization implies also the generalization of statistical theory as the new road for the unification of macroscopic and microscopic complexity.

If $P[\vec{u}(\vec{x},t)]$ is the probability of the entire field path in the field state space of the distributed system, then we can extend the theory of generating function of moments and cumulants for the probabilistic description of the paths [15, 63]. The n-point field correlation functions (n-points moments) can be estimated by using the field path probability distribution and field path (functional) integration:

$$\langle u(\vec{x}_1,t_1)u(\vec{x}_2,t_2)...u(x_n,t_n)\rangle = \int \Delta \vec{u} P[\vec{u}(\vec{x},t)] u(\vec{x}_1,t_1)...u(\vec{x}_n,t_n) \qquad (25)$$

For Gaussian random processes which happen to be near equilibrium the $n$ – th point moments with $n > 2$ are zero, correspond to Markov processes while far from equilibrium it is possible non-Gaussian (with infinite nonzero moments) processes to be developed. According to Haken [15] the characteristic function (or generating function) of the probabilistic description of paths:

$$[u(x,t)] \equiv (u(\vec{x}_1,t_1), u(\vec{x}_2,t_2), ..., u(\vec{x}_n,t_n)) \qquad (26)$$

is given by the relation:

$$\Phi_{path}(j_1(t_1), j_2(t_2), ..., j_n(t_n)) = \left\langle \exp i \sum_{i=1}^{N} j_i u(\vec{x}_i, t_i) \right\rangle_{path} \qquad (27)$$

while the path cumulants $K_s(t_{a_1}...t_{a_s})$ are given by the relations:

$$\Phi_{path}(j_1(t_1), j_2(t_2), ..., j_n(t_n)) = \exp\left\{ \sum_{s=1}^{\infty} \frac{i^s}{s!} \sum_{a_1,...,a_s=1}^{n} K_s(t_{a_1}...t_{a_s}) \cdot j_{a_1}...j_{a_s} \right\} \qquad (28)$$

and the $n$ – point path moments are given by the functional derivatives:

$$\langle u(\vec{x}_1,t_1), u(\vec{x}_2,t_2), ..., u(\vec{x}_n,t_n) \rangle = \left( \delta^n \Phi(\{j_i\}) / \delta j_1...\delta j_n \right) t\{j_i\} = 0 \qquad (29)$$

For Gaussian stochastic field processes the cumulants except the first two vanish $(k_3 = k_4 = ...0)$. For non-Gaussian processes it is possible to be developed long range correlations as the cumulants of higher than two order are non-zero [15]. This is the deeper meaning of non-equilibrium self-organization and ordering of complex systems. The characteristic function of the dynamical stochastic field system is related to the partition functions of its statistical description, while the cumulant development and multipoint moments generation can be related with the BBGKY statistical hierarchy of the statistics as well as with the Feynman diagrams approximation of the stochastic field system [37, 64]. For dynamical systems near equilibrium only the second order cumulants is non-vanishing, while far from equilibrium field fluctuations with higher – order non-vanishing cumulants can be developed.

Finally, we can understand how the non-linear dynamics correspond to self-organized states as the high-order (infinite) non-vanishing cumulants can produce the non-integrability of the dynamics. From this point of view the Boltzmann – Vlassov linear or non-linear instabilities or the linearized MHD models are inefficient to produce the non-Gaussian, holistic (non-local) and self-organized complex character of non-equilibrium space plasmas. That is, far from equilibrium space plasma states can be developed including long range correlations of field and particles with non-Gaussian distributions of their dynamic variables. As we show in the next section such states, as indicated by in situ observations in space plasma, reveal the necessity of new theoretical tools for their understanding much different from the known linear, non-linear instabilities, or reconnection models included in MHD theory and the collisionless Boltzmann – Vlassov approximation [25-28,44].

**2.3 Fractal generalization of space plasma dynamics.**

Fractal integrals and fractal derivatives are related with the fractal contraction transformation of phase space as well as contraction transformation of space time in analogy with the fractal contraction transformation of the Cantor set [51, 52]. Also, the fractal extension of dynamics includes an extension of non-Gaussian scale invariance, related to the multiscale coupling and non-equilibrium extension of the renormalization group theory [20]. Moreover Tarasov [19], Coldfain [23], Cresson [22], El-Nabulsi [21] and other scientists generalized the classical or quantum dynamics in a continuation of the original break through of El-Naschie[17], Nottale[16], Castro[18] and others concerning the fractal generalization of physical theory. According to Tarasov [19] the fundamental theorem of Riemann – Liouville fractional calculus is the generalization of the known integer integral – derivative theorem as follows:

if
$$F(x) = {}_aI_x^a f(x) \tag{30}$$

then
$$_aD_x^a F(x) = f(x) \tag{31}$$

where ${}_aI_x^a$ is the fractional Riemann – Liouville according to:

$$_aI_x^a f(x) \equiv \frac{1}{\Gamma(a)} \int_a^x \frac{f(x')dx'}{(x-x')^{1-a}} \tag{32}$$

and ${}_aD_x^a$ is the Caputo fractional derivative according to:

$$_aD_x^a F(x) = {}_aI_x^{n-a} D_x^n F(x) =$$
$$= \frac{1}{\Gamma(n-a)} \int_a^x \frac{dx'}{(x-x')^{1+a-n}} \frac{d^n F(x)}{dx^n} \tag{33}$$

for $f(x)$ a real valued function defined on a closed interval $[a,b]$.

In the next we summarize the basic concepts of the fractal generalization of mechanics and electromagnetic theory as well as the fractal generalization of Liouville and MHD theory following Tarasov [19]. According to previous descriptions, the far from equilibrium space plasma dynamics includes fractal or multi-fractal distribution of fields and particles, as well as spatial fractal temporal distributions. This state can be described by the fractal generalization of classical theory: Lagrange and Hamilton equations of dynamics, Liouville theory, Fokker Planck equations and Bogoliubov hierarchy equations, Maxwell's equations, hydrodynamics and Magnetohydrodynamics (MHD) equations. In general, the fractal distribution of a physical quantity ($M$) obeys a power law relation:

$$M_D \sim M_0 \left(\frac{R}{R_0}\right)^D \tag{34}$$

where ($M_D$) is the fractal mass of the physical quantity ($M$) in a ball of radius ($R$) and ($D$) is the distribution fractal dimension. For a fractal distribution with local density $\rho(\vec{x})$ the fractal generalization of Euclidean space integration reads as follows:

$$M_D(W) = \int_W \rho(x) dV_D \tag{35}$$

where
$$dV_D = C_3(D, \vec{x}) dV_3 \tag{36}$$

and
$$C_3(D, \vec{x}) = \frac{2^{3-D} \Gamma(3/2)}{\Gamma(D/2)} |\vec{x}|^{D-3} \tag{37}$$

Similarly the fractal generalization of surface and line Euclidean integration is obtained by using the relations:

$$dS_d = C_2(d, \vec{x}) dS_2 \tag{38}$$

$$C_2(d, \vec{x}) = \frac{2^{2-d}}{\Gamma(d/2)} |\vec{x}|^{d-2} \tag{39}$$

for the surface fractal integration and

$$dl_\gamma = C_1(\gamma, \vec{x}) dl_1 \tag{40}$$

$$C_1(\gamma, \vec{x}) = \frac{2^{1-\gamma} \Gamma(1/2)}{\Gamma(\gamma/2)} |\vec{x}|^{\gamma-1} \tag{41}$$

for the line fractal integration. By using the fractal generalization of integration and the corresponding generalized Gauss's and Stoke's theorems we can transform fractal integral laws to fractal and non-local differential laws [19] The fractional generalization of classical dynamics (Hamilton Lagrange and Liouville theory) can be obtained by the fractional generalization of phase space quantative description [19]. For this we use the fractional power of coordinates:

$$X^a = \text{sgn}(x)|x|^a \tag{42}$$

where $\text{sgn}(x)$ is equal to $+1$ for $x \geq 0$ and equal to $-1$ for $x < 0$.

The fractional measure $M_a(B)$ of a $n$-dimension phase space region $(B)$ is given by the equation:

$$M_a(B) = \int_B g(a) d\mu_a(q,p) \tag{43}$$

where $d\mu_a(q,p)$ is a phase space volume element:

$$d\mu_a = \Pi \frac{dq_K^a \wedge dp_K^a}{[a\Gamma(a)]^2} \tag{44}$$

where $g(a)$ is a numerical multiplier and $dq_K^a \wedge dp_K^a$ means the wedge product.

The fractional Hamilton's approach can be obtained by the fractal generalization of the Hamilton action principle:

$$S = \int [pq - H(t,p,q)] dt \tag{45}$$

The fractal Hamilton equations:

$$\left(\frac{dq}{at}\right)^q = \Gamma(2-a) p^{a-1} D_p^a H \tag{46}$$

$$D_t^a p = -D_q^a H \tag{47}$$

while the fractal generalization of the Lagrange's action principle:

$$S = \int L(t,q,u) dt \tag{48}$$

Corresponds to the fractal Lagrange equations:

$$D_q^a L - \Gamma(2-a) D_t^a \left[ D_U^a L \right]_{U=\dot{q}} = 0 \tag{49}$$

Similar fractal generalization can be obtained for dissipative or non-Hamiltonian systems [19]. The fractal generalization of Liouville equation is given also as:

$$\frac{\partial \tilde{p}_N}{\partial t} = L_N \tilde{p}_N \tag{50}$$

where $\tilde{p}_N$ and $L_N$ are the fractal generalization of probability distribution function and the Liouville operator correspondingly. The fractal generalization of Bogoliubov hierarchy can be obtained by using the fractal Liouville equation as well as the fractal Fokker Planck hydrodynamical - magnetohydrodynamical approximations [19].

The fractal generalization of classical dynamical theory for dissipative systems includes the non-Gaussian statistics as the fractal generalization of Boltzmann – Gibbs statistics.

Finally the far from equilibrium statistical mechanics can be obtained by using the fractal extension of the path integral method. The fractional Green function of the dynamics is given by the fractal generalization of the path integral:

$$K_a\left(x_f,t_f;x_i,t_i\right) \simeq \int_{x_i}^{x_f} D[x_a(\tau)]\exp\left[\frac{i}{\hbar}S_a(\gamma)\right]$$
$$\simeq \sum_{\{\gamma\}} \exp\left[\frac{i}{\hbar}S_a(\gamma)\right] \quad (51)$$

where $K_a$ is the probability amplitude (fractal quantum mechanics) or the two point correlation function (statistical mechanics), $D[x_a(\tau)]$ means path integration on the sum $\{\gamma\}$ of fractal paths and $S_a(\gamma)$ is the fractal generalization of the action integral [21]:

$$S_a[\gamma] = \frac{1}{\Gamma(a)} \int_{x_i}^{x_f} L\left(D_\gamma^a q(\tau),\tau\right)(t-\tau)^{a-1} d\tau \quad (52)$$

## 2.4 The Highlights of Tsallis Theory

As we show in the next sections of this study, everywhere in space plasmas we can ascertain the presence of Tsallis statistics. This discovery is the continuation of a more general ascertainment of Tsallis q-extensive statistics from the macroscopic to the microscopic level [1].

In our understanding the Tsallis theory, more than a generalization of thermodynamics for chaotic and complex systems, or a non-equilibrium generalization of B-G statistics, can be considered as a strong theoretical vehicle for the unification of macroscopic and microscopic physical complexity. From this point of view Tsallis statistical theory is the other side of the modern fractal generalization of dynamics while its essence is nothing else than the efficiency of self-organization and development of long range correlations of coherent structures in complex systems.

From a general philosophical aspect, the Tsallis q-extension of statistics can be identified with the activity of an ordering principle in physical reality, which cannot be exhausted with the local interactions in the physical systems, as we noticed in previous sections.

### 2.4.1 The non-extensive entropy ($S_q$).

It was for first time that Tsallis [1], inspired by multifractal analysis, conceived that the Boltzmann – Gibbs entropy:

$$S_{BG} = -K \sum p_i \ln p_i \, , \, i=1,2,...,N \quad (53)$$

is inefficient to describe all the complexity of non-linear dynamical systems. The Boltzmann – Gibbs statistical theory presupposes ergodicity of the underlying dynamics in the system phase space. The complexity of dynamics which is far beyond the simple ergodic complexity, it can be described by Tsallis non-extensive statistics, based on the extended concept of $q$ – entropy:

$$S_q = k\left(1-\sum_{i=1}^{N} p_i^q\right)/(q-1) \quad (54)$$

for discrete state space or

$$S_q = k\left[1-\int [p(x)]^q dx\right]/(q-1) \quad (55)$$

for continuous state space.

For a system of particles and fields with short range correlations inside their immediate neighborhood, the Tsallis $q$ – entropy $S_q$ asymptotically leads to Boltzmann – Gibbs entropy ($S_{BG}$) corresponding to the value of $q = 1$. For probabilistically dependent or correlated systems $A, B$ it can be proven that:

$$S_q(A+B) = S_q(A) + S_q(B/A) + (1-q)S_q(A)S_q(B/A)$$
$$= S_q(B) + S_q(A/B) + (1-q)S_q(B)S_q(A/B) \quad (56)$$

where $S_q(A) \equiv S_q(\{p_i^A\}), S_q(B) \equiv Sq(\{p_i^B\}), S_q(B/A)$ and $S_q(A/B)$ are the conditional entropies of systems $A, B$ [1]. When the systems are probabilistically independent, then relation (3.1.4) is transformed to:

$$S_q(A+B) = S_q(A) + S_q(B) + (1-q)S_q(A)S_q(B) \quad (57)$$

The dependent (independent) property corresponds to the relation:

$$p_{ij}^{A+B} \neq p_i^A p_j^B \left( p_{ij}^{A+B} = p_i^A p_j^B \right) \quad (58)$$

Comparing the Boltzmann – Gibbs ($S_{BG}$) and Tsallis ($S_q$) entropies, we conclude that for non-existence of correlations $S_{BG}$ is extensive whereas $S_q$ for $q \neq 1$ is non-extensive. In contrast, for global correlations, large regions of phase – space remain unoccupied. In this case $S_q$ is non-extensive either $q = 1$ or $q \neq 1$.

**2.4.2 The $q$ – extension of statistics and Thermodynamics.**

Non-linearity can manifest its rich complex dynamics as the system is removed far from equilibrium. The Tsallis $q$ – extension of statistics is indicated by the non-linear differential equation $dy/dx = y^q$. The solution of this equation includes the $q$ – extension of exponential and logarithmic functions:

$$e_q^x = [1 + (1-q)x]^{1/(1-q)} \quad (59)$$
$$\ln_q x = \left(x^{1-q} - 1\right)/(1-q) \quad (60)$$

and

$$p_{opt}(x) = e_q^{-\beta_q[f(x)-F_q]} / \int dx' e_q^{-\beta_q[f(x')-F_q]} \quad (61)$$

for more general $q$ – constraints of the forms $\langle f(x) \rangle_q = F_q$. In this way, Tsallis $q$ – extension of statistical physics opened the road for the $q$ – extension of thermodynamics and general critical dynamical theory as a non-linear system lives far from thermodynamical equilibrium. For the generalization of Boltzmann-Gibbs nonequilibrium statistics to Tsallis nonequilibrium q-statistics we follow Binney [37]. In the next we present q-extended relations, which can describe the non-equilibrium fluctuations and $n$ – point correlation function ($G$) can be obtained by using the Tsallis partition function $Z_q$ of the system as follows:

$$G_q^n(i_1, i_2, ..., i_n) \equiv \langle s_{i_1}, s_{i_2}, ..., s_{i_n} \rangle_q = \frac{1}{z} \frac{\partial^n Z_q}{\partial j_{i_1} \cdot \partial j_{i_2} ... \partial j_{i_n}} \quad (62)$$

Where $\{s_i\}$ are the dynamical variables and $\{j_i\}$ their sources included in the effective – Lagrangian of the system. Correlation (Green) equations (62) describe discrete variables, the $n$ – point correlations for continuous distribution of variables (random fields) are given by the functional derivatives of the functional partition as follows:

$$G_q^n(\vec{x}_1, \vec{x}_2, ..., \vec{x}_n) \equiv \langle \varphi(\vec{x}_1)\varphi(\vec{x}_2)...\varphi(\vec{x}_n) \rangle_q = \frac{1}{Z} \frac{\delta}{\delta J(\vec{x}_1)} ... \frac{\delta}{\delta J(\vec{x}_n)} Z_q(J) \quad (63)$$

where $\varphi(\vec{x})$ are random fields of the system variables and $j(\vec{x})$ their field sources. The connected $n$ – point correlation functions $G_i^n$ are given by:

$$G_q^n(\vec{x}_1, \vec{x}_2, ..., \vec{x}_n) \equiv \frac{\delta}{\delta J(\vec{x}_1)} ... \frac{\delta}{\delta J(\vec{x}_n)} \log Z_q(J) \qquad (64)$$

The connected $n$ – point correlations correspond to correlations that are due to internal interactions defined as [37]:

$$G_q^n(\vec{x}_1, \vec{x}_2, ..., \vec{x}_n) \equiv \langle \varphi(\vec{x}_1)...\varphi(\vec{x}_n) \rangle_q - \langle \varphi(x_1)...\varphi(x_n) \rangle_q \qquad (65)$$

The probability of the microscopic dynamical configurations is given by the general relation:

$$P(conf) = e^{-\beta S_{conf}} \qquad (66)$$

where $\beta = 1/kt$ and $S_{conf}$ is the action of the system, while the partition function $Z$ of the system is given by the relation:

$$Z = \sum_{conf} e^{-\beta S_{conf}} \qquad (67)$$

The $q$ – extension of the above statistical theory can be obtained by the $q$ – partition function $Z_q$. The $q$ – partition function is related with the meta-equilibrium distribution of the canonical ensemble which is given by the relation:

$$p_i = e_q^{-\beta q(E_i - V_q)/Z_q} \qquad (68)$$

with

$$Z_q = \sum_{conf} e_q^{-\beta q(E_i - V_q)} \qquad (69)$$

and

$$\beta_q = \beta / \sum_{conf} p_i^q \qquad (70)$$

where $\beta = 1/KT$ is the Lagrange parameter associated with the energy constraint:

$$\langle E \rangle_q \equiv \sum_{conf} p_i^q E_i / \sum_{conf} p_i^q = U_q \qquad (71)$$

The $q$ – extension of thermodynamics is related with the estimation of $q$ – Free energy ($F_q$) the $q$ – expectation value of internal energy $(U_q)$ the $q$ – specific heat $(C_q)$ by using the $q$ – partition function:

$$F_q \equiv U_q - TS_q = -\frac{1}{\beta} \ln_q Z_q \qquad (72)$$

$$U_q = \frac{\partial}{\partial \beta} \ln_q Z_q, \frac{1}{T} = \frac{\partial S_q}{\partial U_q} \qquad (73)$$

$$C_q \equiv T\frac{\partial \delta_q}{\partial T} = \frac{\partial U_q}{\partial T} = -T\frac{\partial^2 F_q}{\partial T^2} \qquad (74)$$

## 2.4.3 The Tsallis $q$ – extension of statistics via the fractal extension of dynamics.

At the equilibrium thermodynamical state the underlying statistical dynamics is Gaussian ($q=1$). As the system goes far from equilibrium the underlying statistical dynamics becomes non-Gaussian ($q \neq 1$). At the first case the phase space includes ergodic motion corresponding to normal diffusion process with mean-squared jump distances proportional to the time $\langle x^2 \rangle \sim t$ whereas far from equilibrium the phase space motion of the dynamics becomes chaotically self-organized corresponding to anomalous diffusion process with mean-squared jump distances $\langle x^2 \rangle \sim t^a$, with $a<1$ for sub-diffusion and $a>1$ for super-diffusion. The equilibrium normal-diffusion process is described by a chain equation of the Markov-type:

$$W(x_3,t_3;x_1,t_1) = \int dx_2 W(x_3,t_3;x_2,t_2) W(x_2,t_2;x_1,t_1) \quad (75)$$

where $W(x,t;x',t')$ is the probability density for the motion from the dynamical state $(x',t')$ to the state $(x,t)$ of the phase space. The Markov process can be related to a random differential Langevin equation with additive white noise and a corresponding Fokker – Planck probabilistic equation [20] by using the initial condition:

$$\lim_{\Delta t \to 0} W(x,y;\Delta t) = \delta(x-y) \quad (76)$$

This relation means no memory in the Markov process and help to obtain the expansion:

$$W(x,y;\Delta t) = \delta(x-y) + a(y;\Delta t)\delta'(x-y) + \frac{1}{2}b(y;\Delta t)\delta''(x-y) \quad (77)$$

where $A(y;\Delta t)$ and $B(y;\Delta t)$ are the first and second moment of the transfer probability function $W(x,y;\Delta t)$:

$$a(y;\Delta t) = \int dx(x-y)W(x,y;\Delta t) \equiv \langle\langle \Delta y \rangle\rangle \quad (78)$$

$$b(y;\Delta t) = \int dx(x-y)^2 W(x,y;\Delta t) \equiv \langle\langle (\Delta y)^2 \rangle\rangle \quad (79)$$

By using the normalization condition:

$$\int dy W(x,y;\Delta t) = 1 \quad (80)$$

we can obtain the relation:

$$a(y;\Delta t) = -\frac{1}{2}\frac{\partial b(y;\Delta t)}{\partial y} \quad (81)$$

The Fokker – Planck equation which corresponds to the Markov process can be obtained by using the relation:

$$\frac{\partial p(x,t)}{\partial t} = \lim_{\Delta t \to 0} \frac{1}{\Delta t}\left[\int_{-\infty}^{+\infty} dy W(x,y;\Delta t)p(y,t) - p(x,t)\right] \quad (82)$$

where $p(x,t) \equiv W(x,x_0;t)$ is the probability distribution function of the state $(x,t)$ corresponding to large time asymptotic, as follows:

$$\frac{\partial P(x,t)}{\partial t} = -\nabla_x(AP(x,t)) + \frac{1}{2}\nabla_x^2(BP(x,t)) \quad (83)$$

where $A(x)$ is the flow coefficient:

$$A(x,t) \equiv \lim_{\Delta t \to 0} \frac{1}{\Delta t} \langle\langle \Delta x \rangle\rangle \tag{84}$$

and $B(x,t)$ is the diffusion coefficient:

$$B(\vec{x},t) \equiv \lim_{\Delta t \to 0} \frac{1}{\Delta t} \langle\langle \Delta x^2 \rangle\rangle \tag{85}$$

The Markov process is a Gaussian process as the moments $\lim_{\Delta t \to 0} \langle\langle \Delta x^m \rangle\rangle$ for $m > 2$ are zero [63]. The stationary solutions of F-P equation satisfy the extremal condition of Boltzmann – Gibbs entropy:

$$S_{BG} = -K_B \int p(x) \ln p(x) dx \tag{86}$$

corresponding to the known Gaussian distribution:

$$p(x) \sim \exp(-x^2 / 2\sigma^2) \tag{87}$$

According to Zaslavsky [20] the fractal extension of Fokker – Planck (F-P) equation can be produced by the scale invariance principle applied for the phase space of the non-equilibrium dynamics. As it was shown by Zaslavsky for strong chaos the phase space includes self similar structures of islands inside islands dived in the stochastic sea [20]. The fractal extension of the FPK equation (FFPK) can be derived after the application of a Renormalization group of anomalous kinetics (RGK):

$$\hat{R}_K : s' = \lambda_s S, \; t' = \lambda_t t$$

where $s$ is a spatial variable and t is the time.
Correspondingly to the Markov process equations:

$$\frac{\partial^\beta p(\xi,t)}{\partial t^\beta} \equiv \lim_{\Delta t \to 0} \frac{1}{(\Delta t)^\beta} \left[ W(\xi,\xi_0; t + \Delta t) - W(\xi,\xi_0; t) \right] \tag{88}$$

$$W(\xi,n;\Delta t) = \delta(\xi - n) + A(n;\Delta t)\delta^{(\alpha)}(\xi - n) + \frac{1}{2} B(n;\Delta t)\delta^{(2a)}(\xi - n) + ... \tag{89}$$

as the space-time variations of probability W are considered on fractal space-time variables $(t,\xi)$ with dimensions $(\beta,a)$.
For fractal dynamics $a(n;\Delta t)$, $b(n;\Delta t)$ satisfy the equations:

$$a(n;\Delta t) = \int |n - \xi|^\alpha W(\xi,n;\Delta t) d\xi \equiv \langle\langle |\Delta\xi|^a \rangle\rangle \tag{90}$$

$$b(n;\Delta t) = \int |n - \xi|^{2\alpha} W(\xi,n;\Delta t) d\xi \equiv \langle\langle |\Delta\xi|^{2a} \rangle\rangle \tag{91}$$

and the limit equations:

$$A(\xi) = \lim_{\Delta t \to 0} \frac{a(\xi;\Delta t)}{(\Delta t)^\beta} \tag{92}$$

$$B(\xi) = \lim_{\Delta t \to 0} \frac{b(\xi;\Delta t)}{(\Delta t)^\beta} \tag{93}$$

By them we can obtain the FFPK equation.
Far from equilibrium the non-linear dynamics can produce phase space topologies corresponding to various complex attractors of the dynamics. In this case the extended complexity of the dynamics corresponds to the generalized strange kinetic Langevin equation with correlated and multiplicative noise components and extended fractal Fokker – Planck - Kolmogorov equation (FFPK) [20, 65]. The $q$ – extension of statistics by Tsallis can be related with the strange kinetics and the fractal extension of dynamics through the Levy process:

$$P(x_n, t_n; x_0 t_0) = \int dx_1 ... dx_{N-1} P(x_N, t_N; x_{N-1}, t_{N-1}) ... P(x_1, t_1; x_0, t_0) \quad (94)$$

The Levy process can be described by the fractal F-P equation:

$$\frac{\partial^\beta P(x,t)}{\partial t^\beta} = \frac{\partial^a}{\partial(-x)^a}[A(x)P(x,t)] + \frac{\partial^{a+1}}{\partial(-x)^{a+1}}[B(x)P(x,t)] \quad (95)$$

where $\partial^\beta/\partial t^\beta$, $\partial^a/\partial(-x)^a$ and $\partial^{a+1}/\partial(-x)^{a+1}$ are the fractal time and space derivatives correspondingly [20]. The stationary solution of the F F-P equation for large $x$ is the Levy distribution $P(x) \sim x^{-(1+\gamma)}$. The Levy distribution coincides with the Tsallis $q$–extended optimum distribution (3.2.4) for $q = (3+\gamma)/(1+\gamma)$. The fractal extension of dynamics takes into account non-local effects caused by the topological heterogeneity and fractality of the self-organized phase – space. Also the fractal geometry and the complex topology of the phase – space introduce memory in the complex dynamics which can be manifested as creation of long range correlations, while, oppositely, in Markov process we have complete absence of memory.

In general, the fractal extension of dynamics as it was done until now from Zaslavsky, Tarasov and other scientists indicate the internal consistency of Tsallis $q$–statistics as the non-equilibrium extension of B-G statistics with the fractal extension of classical and quantum dynamics. Concerning the space plasmas the fractal character of their dynamics has been indicated also by many scientists. Indicatively, we refer the fractal properties of sunspots and their formation by fractal aggregates as it was shown by Zelenyi and Milovanov [2, 4], the anomalous diffusion and intermittent turbulence of the solar convection and photospheric motion shown by Ruzmakin et al. [5], the multi-fractal and multi-scale character of space plasmas indicated by Lui [11] and Pavlos et al. [9].

Finally we must notice the fact that the fractal extension of dynamics identifies the fractal distribution of a physical magnitude in space and time according to the scaling relation $M(R) \sim R^a$ with the fractional integration as an integration in a fractal space [19]. From this point of view it could be possible to conclude the novel concept that the non-equilibrium $q$–extension of statistics and the fractal extension of dynamics are related with the fractal space and time themselves [16, 24, 65].

## 3. Theoretical expectations for space plasmas through Tsallis statistical theory and fractal dynamics.

Tsallis $q$–statistics as well as the non-equilibrium fractal dynamics indicate the multi-scale, multi-fractal chaotic and holistic dynamics of space plasmas. Before we present experimental verification of the theoretical concepts described in previous studies as concerns space plasmas in this section we summarize the most significant theoretical expectations.

### 3.1 The $q$ – triplet of Tsallis.

The fractal dynamics in a multi-scale and multi-fractal space described by a fractal Fokker – Planck type equation can be related with the $q$–triplet of Tsallis [1, 9, 32, 33]. The Tsallis $q$–triplet includes three $q$–indices ($q_{sen}, q_{sta}, q_{rel}$) which for the Gaussian dynamics satisfy the relation $q_{sen} = q_{stat} = q_{rel} = 1$. For non-equilibrium and non-Gaussian dynamics, they satisfy the relation $q_{sen} < 1 < q_{stat} < q_{rel}$. The $q_{stat}$ corresponds to the optimum probability distribution function of the magnitude $(x)$:

$$p(x) \propto \left[1-(1-q)\beta_{q_{stat}} x^2\right]^{1/1-q_{stat}} \qquad (96)$$

The $q_{sen}$ index described the deviation $(\xi)$ of neighboring trajectories in phase space by the relation:

$$\xi = \left[1 - \frac{\lambda q_{sen}}{\lambda_1} + \frac{\lambda q_{sen}}{\lambda_1} e^{(1-q_{sen})\lambda_1 t}\right]^{\frac{1}{1-q}} \qquad (97)$$

as well as the multifractal profile of the phase space by the relation:

$$\frac{1}{q_{sen}} = \frac{1}{a_{min}} - \frac{1}{a_{max}} \qquad (98)$$

where $a_{min}(a_{max})$ corresponds to the zero points of the multifractal exponent spectrum $f(a)$ [1, 66].

The $q_{rel}$ corresponds to the relaxation process:

$$\frac{d\Omega}{dt} = -\frac{1}{T_{rel}} \Omega^{q_{rel}} \qquad (99)$$

of a macroscopic observable according to the $q$ – extension of fluctuation – dissipation theory [1].

## 3.2 Multifractal analysis and intermittent turbulence

According to Arimitsu [66] the exponent's spectrum $f(a)$ can be concluded by extremizing the Tsallis entropy functional $S_q$ related to the probability distribution $P(a)$ of the singularities $(a)$ related to the multi-fractal (intermittent) character of the plasmas turbulent dynamics [9]:

$$f(a) = D_0 + \log_2[1-(1-q)\frac{(a-a_o)^2}{2X/\ln 2}]/(1-q)^{-1} \qquad (100)$$

where $a_0$ corresponds to the $q$ – expectation (mean) value of $a$ through the relation:

$$<(a-a_0)^2>_q = (\int da P_q(a)(a-a_0)^2)/\int da P_q(a). \qquad (101)$$

while the $q$ – expectation value $a_0$ corresponds to the maximum of the function $f(a)$ as $df(a)/da|_{a_0} = 0$, with $f(a)$ is related the Renyi fractal dimensions spectrum $D_{\bar{q}}$ according to the relation:

$$T(\bar{q}) = (\bar{q}-1)D_{\bar{q}} \qquad (102)$$

as well as the spectrum $J(p)$ of the structure functions which was first introduced by Kolmogorov [67] as follows:

$$S_p(\ln) = \left\langle |\delta u_n|^p \right\rangle \qquad (103)$$

where $(\delta u_n)^3 = \varepsilon_n$ corresponds to the $n-th$ scale turbulent field increments and is related to the energy dissipation through the relation $(\delta u_n)^3 = \varepsilon_n l_n$ where $l_n = l_0 \delta_n$ is the $n-th$ scale and $\varepsilon_n$ is the energy dissipation in the $n-th$ scale. The $J(p)$ spectrum can be concluded [66] after Tsallis entropy extremization as follows:

$$J(p) = \frac{a_0 p}{3} - \frac{2Xp^2}{q(1+\sqrt{C_{p/3}})} - \frac{1}{1-q}[1-\log_2(1+\sqrt{C_{p/3}})] \qquad (104)$$

The first term $a_0 p/3$ corresponds to the original structure function exponent spectrum given by Kolmogorov theory [67]. According to Kolmogorov's first theory the dissipation of field energy $\varepsilon_n$ is identified with the mean value $\varepsilon_0$ according to the Gaussian self-similar homogeneous turbulence dissipation concept, while $a_0 = 1$ according to multifractal analysis for homogeneous turbulence where the multifractal spectrum consists of a single point. The next terms after the first in the relation (104) correspond to the multifractal structure of intermittence turbulence indicating that the turbulent state is not homogeneous across spatial scales. That is, there is a greater spatial concentration of turbulent activity at smaller than at larger scales. According to Abramenko [8] the intermittent multifractal (inhomogeneous) turbulence is indicated by the general scaling exponent $J(p)$ of the structure functions according to the relation:

$$J(p) = \frac{p}{3} + T^{(u)}(p) + T^{(F)}(p), \qquad (105)$$

where the $T^{(u)}(p)$ term is related with the dissipation of kinetic energy and the $T^{(F)}(p)$ term is related to other forms of field's energy dissipation as the magnetic energy at MHD turbulence. In this study, we present the estimation of $q_{stat}$ index for various space plasma systems. The estimation of all the possible experimental verifications of Tsallis theory will be presented elsewhere according to the description by Pavlos [9, 2011, 2012].

### 3.3 Fractal particle acceleration and fractal energy dissipation

The problem of magnetic energy dissipation and bursty acceleration processes at flares, magnetospheric plasma sheet and other regions of space plasmas is an old and yet resisting problem of space plasma science. Chaotic motion of particles and acceleration through chaotic tear mode instabilities has indicated by Zelenyi [2] or Buchner and Zelenyi [28]. Also turbulent rear mode reconnection and magnetic energy dissipation was indicated by Strauss [25] while the reconnection concept was extended by Chang [3], Consolini [7], Sharma [6]. Fractional kinetics was developed also by Zelenyi [2] and Milovanov [4] concerning the problems of cosmic electrodynamics.

According to Zaslavsky [20], as well as Shlesinger et al. [65] the fractal and multi-scale character of phase space creates new possibilities for the acceleration of charged particles and the magnetic energy dissipation.

The acceleration of energetic particles in spare plasma is caused by inductive electric fields caused by the variation change of the magnetic flux according to faraday law:

$$\oint_c \vec{E} \cdot d\vec{l} = -\frac{d}{dt}\int_s \vec{B} \cdot d\vec{s} \qquad (105)$$

The magnetic field – magnetic flux changes can be related to the turbulent magnetic field diffusion which happens according to MHD equation:

$$\frac{\partial \vec{B}}{\partial t} = \vec{\nabla} X\left(\vec{V} X \vec{B}\right) + n\nabla^2 \vec{B} \qquad (106)$$

However normal Gaussian diffusion process described by Fokker – Planck equation is unable to explain the bursty character of energetic particle acceleration following the bursty development of inductive electric fields after turbulent magnetic flux change [68]. However the fractal extension of dynamics and Tsallis extension of statistics indicate the possibility for a mechanism of fractal dissipation and fractal acceleration process in space plasmas.
According to Tsallis statistics and fractal dynamics the super-diffusion process:

$$\langle R^2 \rangle \sim t^\gamma \quad (107)$$

with $\gamma > 1$ ($\gamma = 1$ for normal diffusion) can be developed in space plasmas. Such process is known as intermittent turbulence or as anomalous diffusion which can be caused by Levy flight process included in fractal dynamics and fractal Fokker – Planck Kolmogorov equation (FFPK). The solution of FFPK equation [20] corresponds to double (temporal, spatial) fractal characteristic function:

$$P(k,t) = \exp\left(-const \cdot t^\beta |k| a\right) \quad (108)$$

Where $P(k,t)$ is the Fourier transform of asymptotic distribution function:

$$P(\xi,t) \sim const \cdot t^\beta / \xi^{1+\alpha}, \ (\xi \to \infty) \quad (109)$$

This distribution is scale invariant with mean displacement:

$$\langle |\xi|^\alpha \rangle \simeq const \cdot t^\beta, \ (t \to \infty) \quad (110)$$

According to this description, the flights of multi-scale and multi-fractal profile can explain the bursty character of magnetic energy dissipation and the bursty character of induced electric fields and charged particle acceleration in space plasmas. The fractal motion of charged particles across the fractal and intermittent topologies of magnetic – electric fields is the essence of strange kinetic [20, 65]. Strange kinetics permits the development of local sources with spatial fractal – intermittent condensation of induced electric fields parallel with fractal – intermittent dissipation of magnetic field energy. These regions correspond to fractal or strange particle accelerators. Such kinds of strange accelerators can be understood by using the Zaslavsky studies for Hamiltonian chaos in anomalous multi-fractal and multi-scale topologies of phase space [20]. Anomalous topology of phase space and fractional Hamiltonian dynamics correspond to dissipative non-Hamiltonian dynamics in the usual phase space [19]. The most important character of fractal kinetics is the wandering of the dynamical state through the gaps of cantori which creates effective barriers for diffusion and long range Levy flights in trapping regions of phase space. Similar Levy flights processes can be developed by the fractal dynamics and intermittent turbulence of the magnetic field which is caused as a consequence of the Faraday law because of long range Levy flights of strong induced electric fields.
 In this theoretical framework it is expected the existence of Tsallis statistics for the distributions of particles and fields coupled with intermittent turbulence characteristics of plasma particles and fields. The fractal dynamics of particles and fields and the corresponding non-extensive Tsallis $q$ – statistical character of the probability distributions indicate the development of a self-organized and globally correlated system of active regions in the plasma systems. This character can be related with deterministic low dimensional chaotic profile of the active regions. This character can be shown by energetic particle observations.

## 4. Comparison of theory with the observations

### 4.1 The Tsallis q-statistics

The traditional scientific point of view is the priority of dynamics over statistics. That is dynamics creates statistics. However for complex system their holistic behaviour does not permit easily such a simplification and division of dynamics and statistics. Tsallis $q$ – statistics and fractal or strange kinetics are two faces of the same complex and holistic (non-reductionist) reality. As Tsallis statistics is an extension of B-G statistics, we can support that the thermic and the dynamical character of a complex system is the manifestation of the same physical process which creates extremized thermic states (extremization of Tsallis entropy), as well as dynamically ordered states. From this point of view the Feynman path integral formulation of physical theory [69] indicates the indivisible thermic and dynamical character of physical reality. After this general investment in the following, we present evidence of Tsallis non-extensive $q$ – statistics for space plasmas. The Tsallis statistics in relation with fractal and chaotic dynamics of space plasmas will be presented in a short coming series of publications.

In next sections we present estimations of Tsallis statistics for various kinds of space plasma's systems. The $q_{stat}$ Tsallis index was estimated by using the observed Probability Distribution Functions (PDF) according to the Tsallis q-exponential distribution:

$$PDF[\Delta Z] \equiv A_q \left[1 + (q-1)\beta_q (\Delta Z)^2\right]^{\frac{1}{1-q}}, \qquad (111)$$

where the coefficients $A_q$, $\beta_q$ denote the normalization constants and $q \equiv q_{stat}$ is the entropic or non-extensivity factor ($q_{stat} \leq 3$) related to the size of the tail in the distributions. Our statistical analysis is based on the algorithm described in [33]. We construct the $PDF[\Delta Z]$ which is associated to the first difference $\Delta Z = Z_{n+1} - Z_n$ of the experimental sunspot time series, while the $\Delta Z$ range is subdivided into little ``cells'' (data binning process) of width $\delta z$, centered at $z_i$ so that one can assess the frequency of $\Delta z$-values that fall within each cell/bin. The selection of the cell-size $\delta z$ is a crucial step of the algorithmic process and its equivalent to solving the binning problem: a proper initialization of the bins/cells can speed up the statistical analysis of the data set and lead to a convergence of the algorithmic process towards the exact solution. The resultant histogram is being properly normalized and the estimated q-value corresponds to the best linear fitting to the graph $\ln_q(p(z_i))$ vs $z_i^2$. Our algorithm estimates for each $\delta_q = 0,01$ step the linear adjustment on the graph under scrutiny (in this case the $\ln_q(p(z_i))$ vs $z_i^2$ graph) by evaluating the associated correlation coefficient *(CC)*, while the best linear fit is considered to be the one maximizing the correlation coefficient. The obtained $q_{stat}$, corresponding to the best linear adjustment is then being used to compute the following equation:

$$G_q(\beta, z) = \frac{\sqrt{\beta}}{C_q} e_q^{-\beta z^2} \qquad (112)$$

where $C_q = \sqrt{\pi} \cdot \Gamma(\frac{3-q}{2(q-1)}) / \sqrt{q-1} \cdot \Gamma(\frac{1}{q-1})$, $1 < q < 3$ for different $\beta$-values. Moreover, we select the $\beta$-value minimizing the $\sum_i [G_{q_{sstat}}(\beta, z_i) - p(z_i)]^2$, as proposed again in [33].

In the following we present the estimation of Tsallis statistics $q_{stat}$ for various cases of space plasma system. Especially, we study the $q$ – statistics for the following space plasma complex systems: I Magnetospheric system, II Solar Wind (magnetic cloud), III Solar activity, IV Cosmic stars, IIV Cosmic Rays.

**4.2 Magnetospheric MHD plasma process.**

For the study of q-statistics we used measurements from spacecraft Geotail in magnetospheric system of calm and sub-storm period. The measurements consist from Vx and Vy componenets in plasma velocity, Ey electric field component, Bz magnetic field component, and elementary particles from electrons and protons. The time zones for calm period are 03:00 UT to 18:30 UT of 7/2/1997 and 01:00 UT to 08:30 UT of 8/2/1997 and for the storm period are 12:00 UT to 21:00 UT of 8/2/1997 and from 12:00 UT of 9/2/1997 to 12:00 UT of 10/2/1997.

During magnetic sub-storm periods in the earth plasma sheet we can observe strong bulk plasma flows as well as burst of energetic particles. These phenomena are considered to follow the development of magnetic reconnection and magnetic energy dissipation in the plasma sheet region. Fig.1 presents in situ in magnetospheric plasma sheet spacecraft observations during the calm periods from 03:00 UT to 18:30 UT of 7/2/1997 and 01:00 UT to 08:30 UT of 8/2/1997 and during the storm periods from 12:00 UT to 21:00 UT of 8/2/1997 and from 12:00 UT of 9/2/1997 to 12:00 UT of 10/2/1997. As the AE index (Fig.1a,b) indicates, the calm magnetospheric period is followed by a strong sub-storm period. The $B_z$ and $V_x$ measurements indicate strong turbulent state of the magnetized plasma in the earth magnetotail. Strong tailward bulk plasma flows ($V_x < 0$) are observed simultaneously with southward magnetic field ($B_z < 0$) component as well as strong earthward bulk plasma flows ($V_x > 0$) simultaneously with northward ($B_z > 0$) magnetic field. During the sub-storm event we can observe strong bulk plasma flows with velocity ~500-1500 km/sec (Fig.1c,e), as well as strong magnetic field change with North – South ($B_Z$) components data values $\pm 20$nT (Fig.1g). At the same period the magnetic field and plasma turbulence causes the induction of strong electric fields with values ~30mV/m (Fig.1d). Generally, the induced electric fields in the plasma sheet can accelerate plasma electrons or protons at energies ~1-2 MeV.

The estimation of $V_x, V_y, B_z$ Tsallis statistics during the substorm period is presented in fig.2(a-i). Fig. 2(a,d,g) shows the experimental time series corresponding to spacecraft observations of bulk plasma flows $V_x, V_y$ and magnetic field $B_z$ component. Fig. 2(b,e,h) presents the q-Gaussian functions $G_q$ described in section 4.1, corresponding to $V_x, V_y, B_z$ times series. The q-Gaussian function presents the best fitting of the experimental distribution function $P(z)$ estimated for the value $q_{stat} = 1.98 \pm 0.06$ for the $V_x$ plasma velocity time series, $q_{stat} = 1.97 \pm 0.05$ for the $V_y$ plasma velocity time series and $q_{stat} = 2.05 \pm 0.04$ for the magnetic field $B_z$ component time series. The q-values of the q-statistics of the signals $V_x, V_y, B_z$ were estimated by the linear correlation fitting between $\ln_q P(z_i)$ and $(z_i)^2$, shown in fig. 2(c,f,i), were $P(z)$ corresponds to the experimental distribution functions, according to the description in section 4.1. The fact that the magnetic field and plasma flow observations obey to non-extensive Tsallis with $q$–values much higher than the Gaussian case ($q = 1$) permit to conclude for magnetospheric plasma the existence of non-equilibrium MHD anomalous diffusion process.

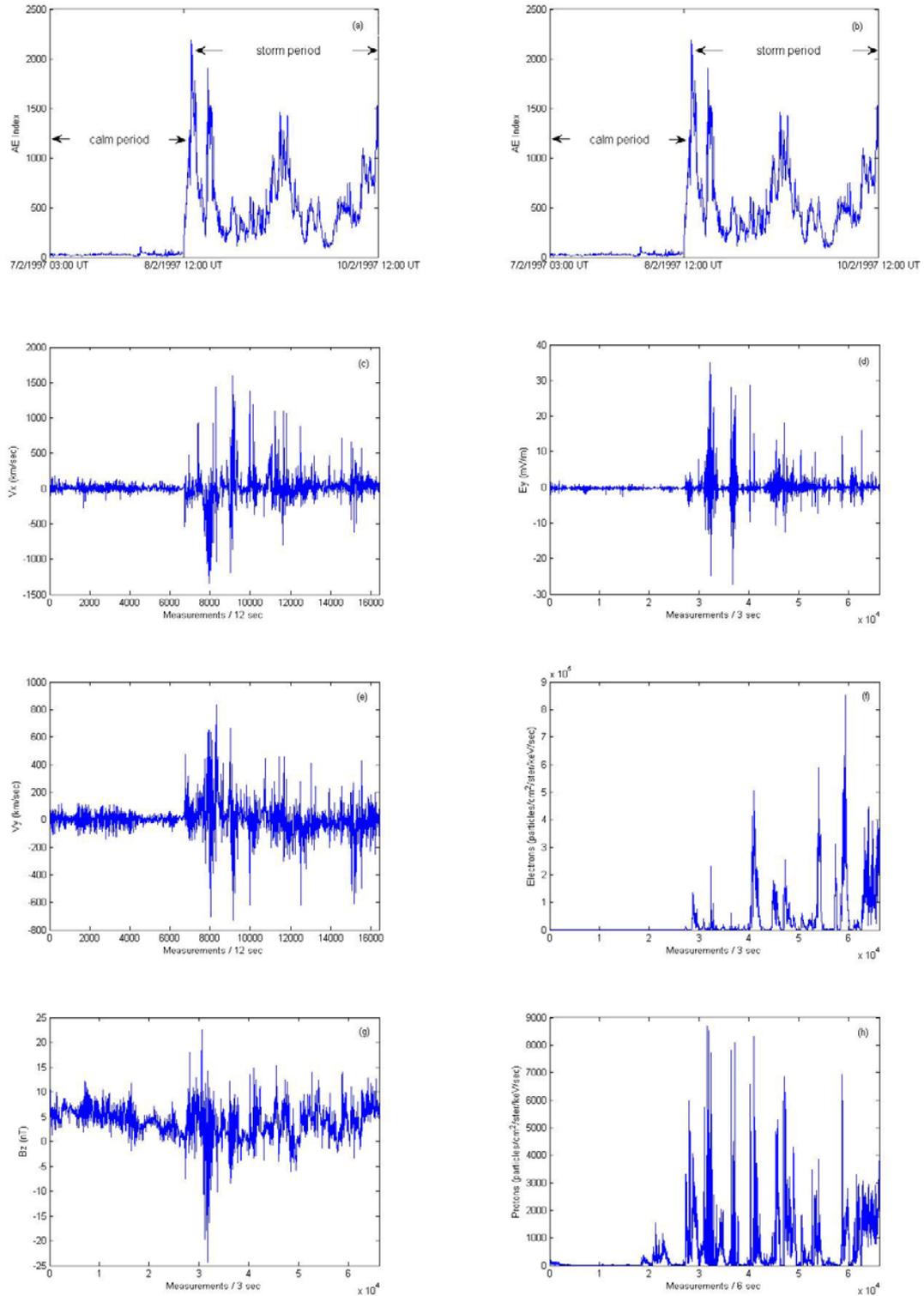

***Figure 1:*** *(**a,b**) AE Index time series with quiet and storm period (**c**) Vx time series with quiet and storm period (**d**) Ey time series with quiet and storm period (**e**) Vy time series with quiet and storm period (**f**) electrons time series with quiet and storm period (**g**) Bz time series with quiet and storm period (**h**) protons time series with quiet and storm period. The calm period corresponds to 03:00-18:30 UT of 7-2-1997 and 01:00-08:30 UT of 8-2-1997 and the storm period to 12:00-21:00 UT of 8-2-1997 and 12:00 UT of 9-2-1997 until 12:00 UT of 10-2-1997.*

### 4.3 Acceleration of energetic particles in the magnetospheric plasma sheet.

Already Tsallis theory has been used for the study of magnetospheric energetic particles non-Gaussian by Voros [70] and Leubner [71]. In the following we study the q-statistics of magnetospheric energetic particle during a strong sub-storm period presented as it was presented previously in Fig.1. We used the data set from the GEOTAIL/EPIC experiment during the period from 12:00 UT to 21:00 UT of 8/2/1997 and from 12:00 UT of 9/2/1997 to 12:00 UT of 10/2/1997. The Tsallis statistics estimated for the magnetospheric electric field and the magnetospheric particles $(e^-, p^+)$ during the storm period is shown in Fig. 3(a-i). Fig. 3(a,d,g) present the spacecraft observations of the magnetospheric electric field $E_y$ component and the magnetospheric electrons $(e^-)$ and protons $(p^+)$. The corresponding Tsallis q-statistics was found to correspond to the q-values: $q_{stat} = 2.49 \pm 0.07$ for the $E_y$ electric field component, $q_{stat} = 2.15 \pm 0.07$ for the energetic electrons and $q_{stat} = 2.49 \pm 0.05$ for the energetic protons. These values reveal clearly non-Gaussian dynamics for the mechanism of electric field development and electrons-protons acceleration during the magnetospheric storm period.

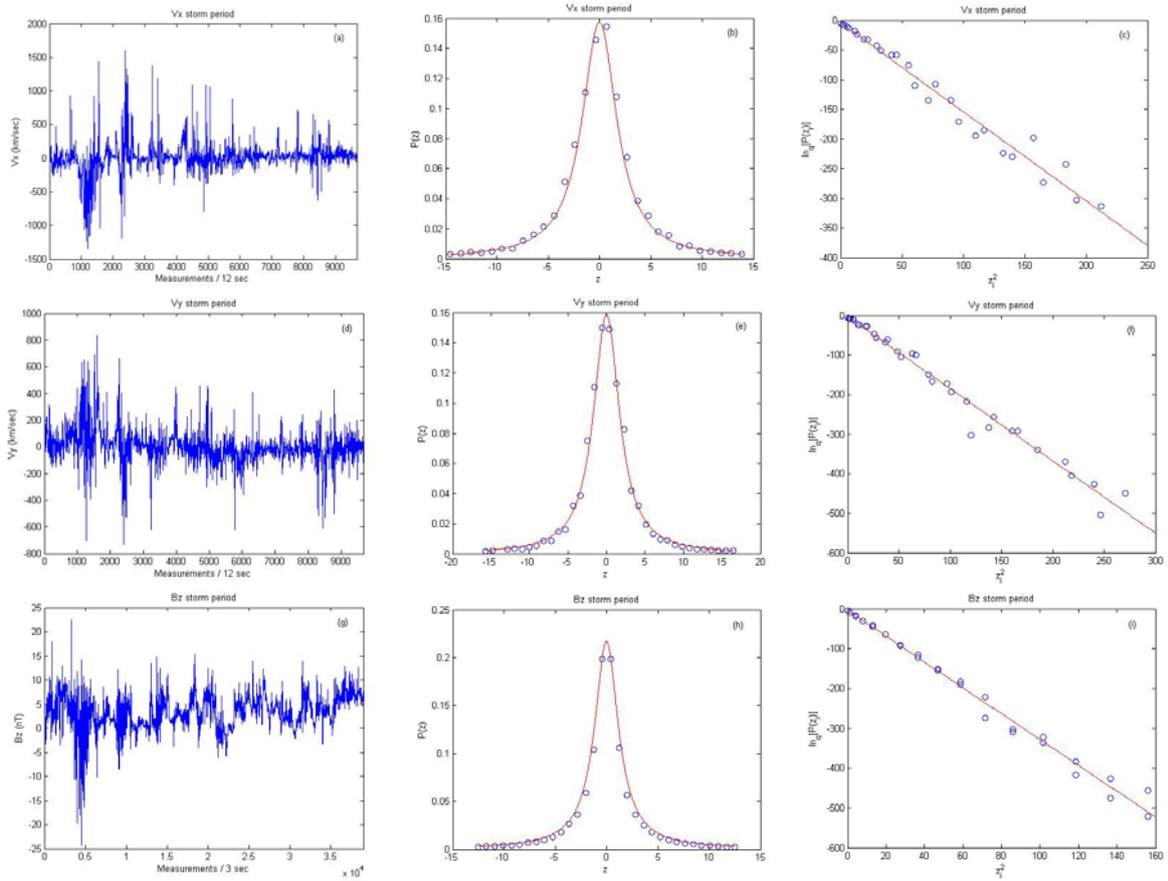

*Figure 2:* (*a*) *Time series of Vx storm period* (*b*) *PDF P(z$_i$) vs. z$_i$ q Guassian function that fits P(z$_i$) for the Vx storm period* (*c*) *Linear Correlation between ln$_q$P(z$_i$) and (z$_i$)$^2$ where q = 1.98 ± 0.06 for the Vx storm period* (*d*) *Time series of Vy storm period* (*e*) *PDF P(z$_i$) vs. z$_i$ q Gaussian function that fits P(z$_i$) for the Vy storm period* (*f*) *Linear Correlation between ln$_q$P(z$_i$) and (z$_i$)$^2$ where q = 1.97 ± 0.05 for the Vy storm period time series* (*g*) *Time series of Bz storm period* (*h*) *PDF P(z$_i$) vs. z$_i$ q Gaussian function that fits P(z$_i$) for the Bz storm period* (*i*) *Linear Correlation between ln$_q$P(z$_i$) and (z$_i$)$^2$ where q = 2.05 ± 0.04 for the Bz storm period.*

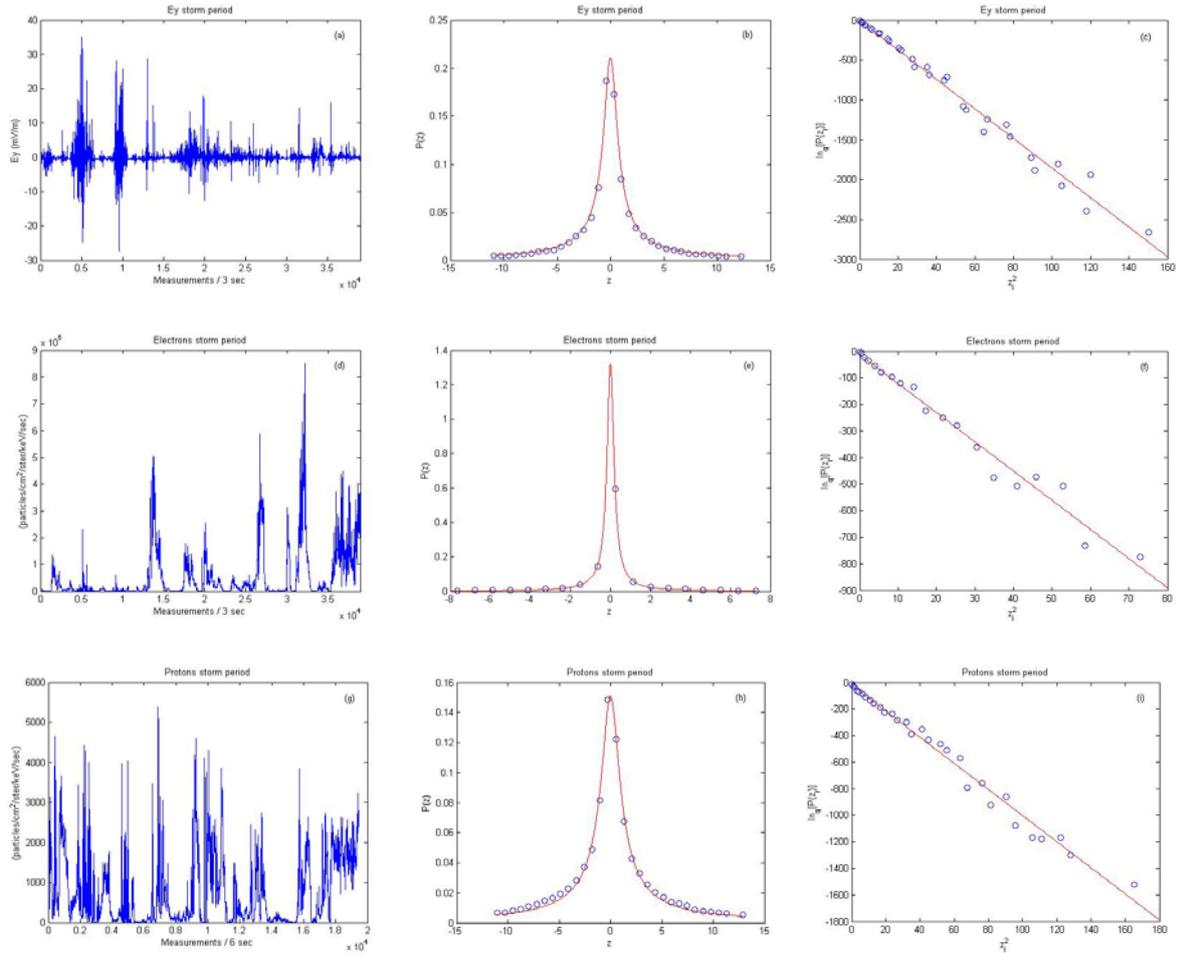

***Figure 3:*** *(**a**) Time series of Ey storm period (**b**) PDF $P(z_i)$ vs. $z_i$ q Guassian function that fits $P(z_i)$ for the Ey storm period (**c**) Linear Correlation between $ln_q P(z_i)$ and $(z_i)^2$ where $q = 2.49 \pm 0.07$ for the Ey storm period (**d**) Time series of electrons storm period (**e**) PDF $P(z_i)$ vs. $z_i$ q Gaussian function that fits $P(z_i)$ for the electrons storm period (**f**) Linear Correlation between $ln_q P(z_i)$ and $(z_i)^2$ where $q = 2.15 \pm 0.07$ for the electrons storm period time series (**g**) Time series of protons storm period (**h**) PDF $P(z_i)$ vs. $z_i$ q Gaussian function that fits $P(z_i)$ for the protons storm period (**i**) Linear Correlation between $ln_q P(z_i)$ and $(z_i)^2$ where $q = 2.49 \pm 0.05$ for the protons storm period.*

### 4.4 Solar activity

The nonlinear dynamics for solar space plasmas and the non-Gaussian space plasma turbulence has been studied by Rempel [72], Carbone et al. [56], de Wit [73], Burlaga [32], Voros [70] and others. Solar activity has as origin the inner core energy production and the radiative zone energy transfer. The turbulent convection zone dynamo process and the magnetic field production create is the outer solar activity which is manifested at the photosphere as sunspot dynamics and coronal loops (prominences) as well as flare bursts, related to magnetic reconnection events and acceleration of charged particles. Coronal mass ejection and solar flares originate in magnetically active regions around visible sunspot groupings. Finally, solar wind is the solar plasma flow in the outer solar corona. In the following we present the q-statistics of Tsallis estimated for solar plasma starting from outer solar corona till the solar convection zone.

### 4.4.1 q-statistics in magnetic clouds at distance 1 AU from the Sun

From the spacecraft ACE, magnetic field experiment (MAG) we take raw data consist from Bx, By, Bz magnetic field components with sampling rate 3 sec. Tha data correspond to substorm period with time zone from 07:27 UT, 20/11/2001 until 03:00 UT, 21/11/2003.

Magnetic clouds are a possible manifestation of a Coronal Mass Ejection (CME) and they represent on third of ejectra observed by satellites. Magnetic cloud behave like a magnetosphere moving through the solar wind. Carbone et al. [56], de Wit [73] estimated non-Gaussian turbulence profile of solar wind. Bourlaga and Vinas [32] estimated the q-statistics of solar wind at the q-value $q_{stat} = 1.75 \pm 0.06$. Fig. 4 presents the q-statistics estimated in the magnetic cloud solar plasma for the three components $B_x, B_y, B_z$ of the magnetic field. The $B_x, B_y, B_z$ time series are shown in Fig. 4(a,d,g). The q-statistics for $B_x, B_y, B_z$ components is shown at Fig. [4(b,c), 4(e,f), 4(h,i)] correspondingly while the q-values were found to be $q_{stat} = 2.07 \pm 0.06$ for the $B_x$, $q_{stat} = 2.00 \pm 0.04$ for the $B_y$ and $q_{stat} = 2.02 \pm 0.04$ for the $B_z$ time series. These values are higher than the values $q_{stat} = 1.75$ estimated from Bourlaga and Vinas [32] at 40 AU.

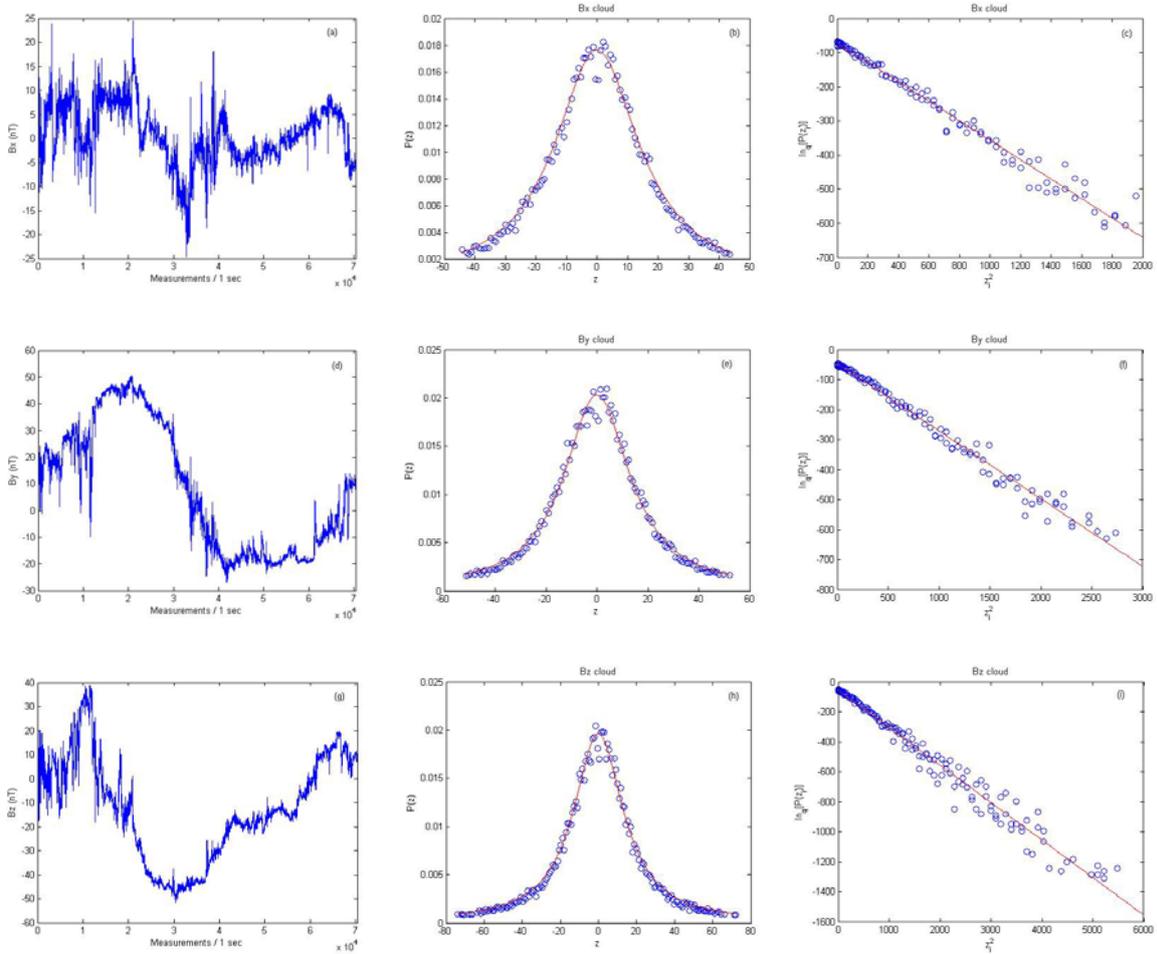

***Figure 4:*** *(**a**) Time series of Bx cloud (**b**) PDF P(z$_i$) vs. z$_i$ q Guassian function that fits P(z$_i$) for the Bx cloud (**c**) Linear Correlation between ln$_q$P(z$_i$) and (z$_i$)$^2$ where q = 2.07 ± 0.06 for the Bx cloud (**d**) Time series of By cloud (**e**) PDF P(z$_i$) vs. z$_i$ q Gaussian function that fits P(z$_i$) for the By cloud (**f**) Linear Correlation between ln$_q$P(z$_i$) and (z$_i$)$^2$ where q = 2.00 ± 0.04 for the By cloud (**g**) Time series of Bz cloud (**h**) PDF P(z$_i$) vs. z$_i$ q Gaussian function that fits P(z$_i$) for the Bz cloud (**i**) Linear Correlation between ln$_q$P(z$_i$) and (z$_i$)$^2$ where q = 2.02 ± 0.04 for the Bz cloud.*

**4.4.2 q-statistics of solar flares and sunspot dynamics**

In this sub-section we present the q-statistics of the sunspot and solar flares complex systems by using data of Wolf number and daily Flare Index. Especially, we use the Wolf number, known as the international sunspot number measures the number of sunspots and group of sunspots on the surface of the sun computed by the formula: (10)$R=k*(10g+s)$ where: $s$ is the number of individual spots, $g$ is the number of sunspot groups and $k$ is a factor that varies with location known as the observatory factor. We analyse a period of 184 years. Moreover we analyse the daily Flare Index of the solar activity that was determined using the final grouped solar flares obtained by NGDC (National Geophysical Data Center). It is calculated for each flare using the formula: $Q=(i*t)$, where "$i$" is the importance coefficient of the flare and "$t$" is the duration of the flare in minutes. To obtain final daily values, the daily sums of the index for the total surface are divided by the total time of observation of that day. The data covers time period from 1/1/1996 to 31/12/2007.

Although solar flares dynamics is coupled to the sunspot dynamics. Karakatsanis and Pavlos [35] and Karakatsanis et al. [36] have shown that the dynamics of solar flares can be discriminated from the sunspot dynamics. Fig. 5 presents the estimation of q-statistics of sunspot index shown in fig. 5(b,c) and the q-statistics of solar flares signal shown in fig. 5(e,g). The q-values for the sunspot index and the solar flares time series were found to be $q_{stat}=1.53\pm0.04$ and $q_{stat}=1.90\pm0.05$ correspondingly. We clearly observe non-Gaussian statistics for both cases but the non-Gaussianity of solar flares was found much stronger than the sunspot index.

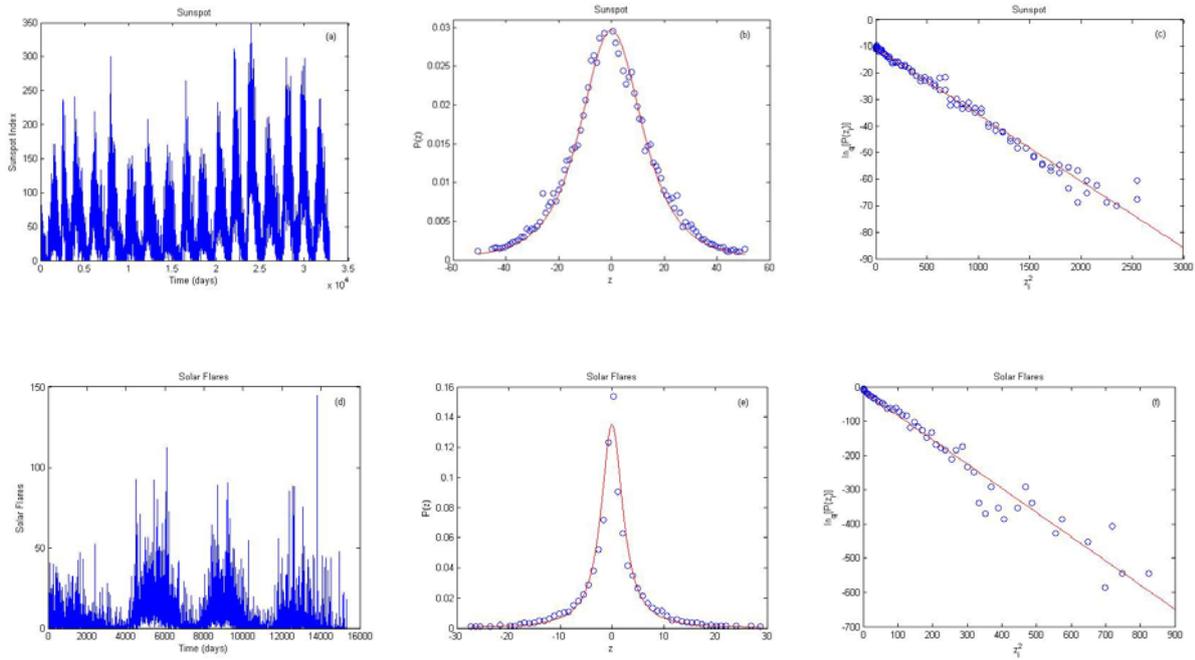

*Figure 5*: (*a*) *Time series of Sunspot Index concerning the period of 184 years* (*b*) *PDF P(z$_i$) vs. z$_i$ q Guassian function that fits P(z$_i$) for the Sunspot Index* (*c*) *Linear Correlation between ln$_q$P(z$_i$) and (z$_i$)$^2$ where q = 1.53 ± 0.04 for the Sunspot Index* (*d*) *Time series of Solar Flares concerning the period of 184 years* (*e*) *PDF P(z$_i$) vs. z$_i$ q Guassian function that fits P(z$_i$) for the Solar Flares* (*f*) *Linear Correlation between ln$_q$P(z$_i$) and (z$_i$)$^2$ where q = 1.90 ± 0.05 for the Solar Flares.*

### 4.4.3 q-statistics of solar energetic electrons and protons

At solar flare regions the dissipated magnetic energy creates strong electric fields according to the theoretical concepts presented in section 3.3. The bursty character of the electric field creates burst of solar energetic particles through a mechanism of solar flare fractal acceleration. According to theoretical concept presented in previous section the fractal acceleration of energetic particles can be concluded by the Tsallis q-extenstion of statistics for non-equilibrium complex states. In the following we present significants verification of this theoretical prediction of Tsallis theory by study the q-statistics of energetic particle acceleration. Finally we analyze energetic particles from spacecraft ACE – experiment EPAM and time zone 1997 day 226 to 2006 day 178 and protons (0.5 – 4) MeV with period 20/6/1986 – 31/5/2006, spacecraft GOES, hourly averaged data. Figure 6 presents the estimation of the solar protons - electrons q-statistics. The q-values for solar energetic protons and electrons time series were found to be $q_{stat} = 2.31 \pm 0.13$ and $q_{stat} = 2.13 \pm 0.06$ correspondingly. Also in this case we clearly observe non-Gaussian statistics for both cases.

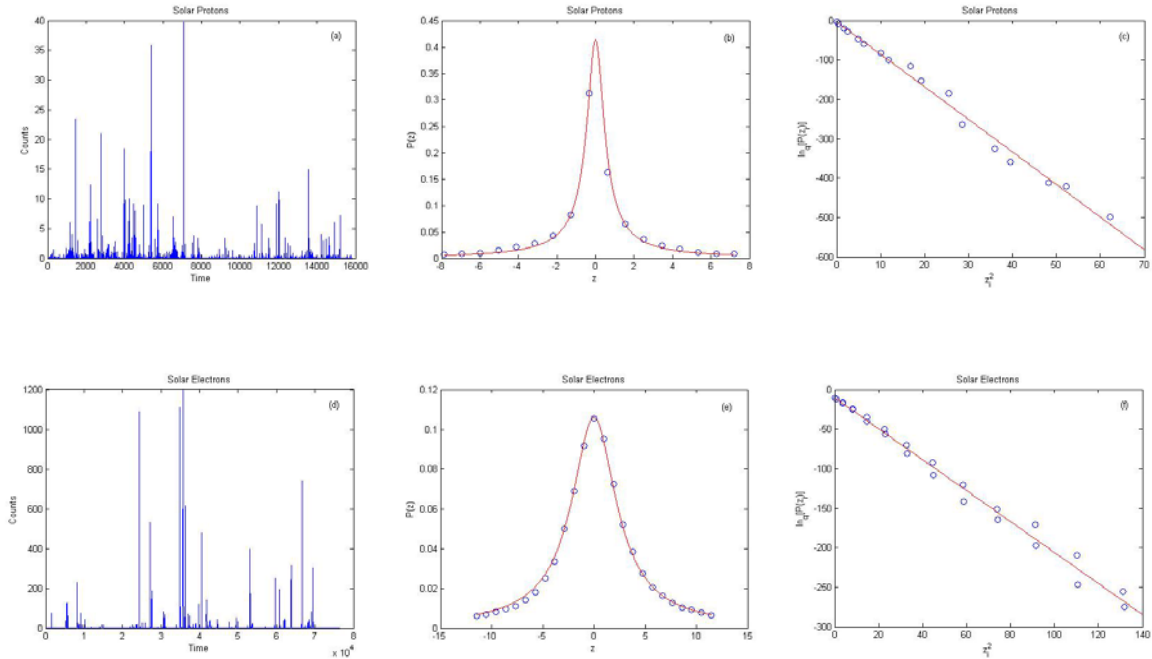

*Figure 6: (a)* Time series of Solar proton *(b)* PDF $P(z_i)$ vs. $z_i$ q Guassian function that fits $P(z_i)$ for the Solar proton data *(c)* Linear Correlation between $ln_q P(z_i)$ and $(z_i)^2$ where q = 2.31 ± 0.13 for the Solar proton *(d)* Time series of Solar electrons *(e)* PDF $P(z_i)$ vs. $z_i$ q Guassian function that fits $P(z_i)$ for the Solar electrons *(f)* Linear Correlation between $ln_q P(z_i)$ and $(z_i)^2$ where q = 2.13 ± 0.06 for the Solar electrons.

### 4.4.4 q-statistics of solar ions ($C^+, Fe^+, O^+$)

In this sub-section we continue the study of the q-statistics for the $C^+, Fe^+, O^+$ solar ions is shown in Fig.[7]. The corresponding q-values were found to be $q_{stat} = 2.27 \pm 0.05$ for the C ion, $q_{stat} = 2.26 \pm 0.03$ for the Fe ions and $q_{stat} = 2.14 \pm 0.06$ for the O time series. We clearly observe non-Gaussian statistics for all cases.

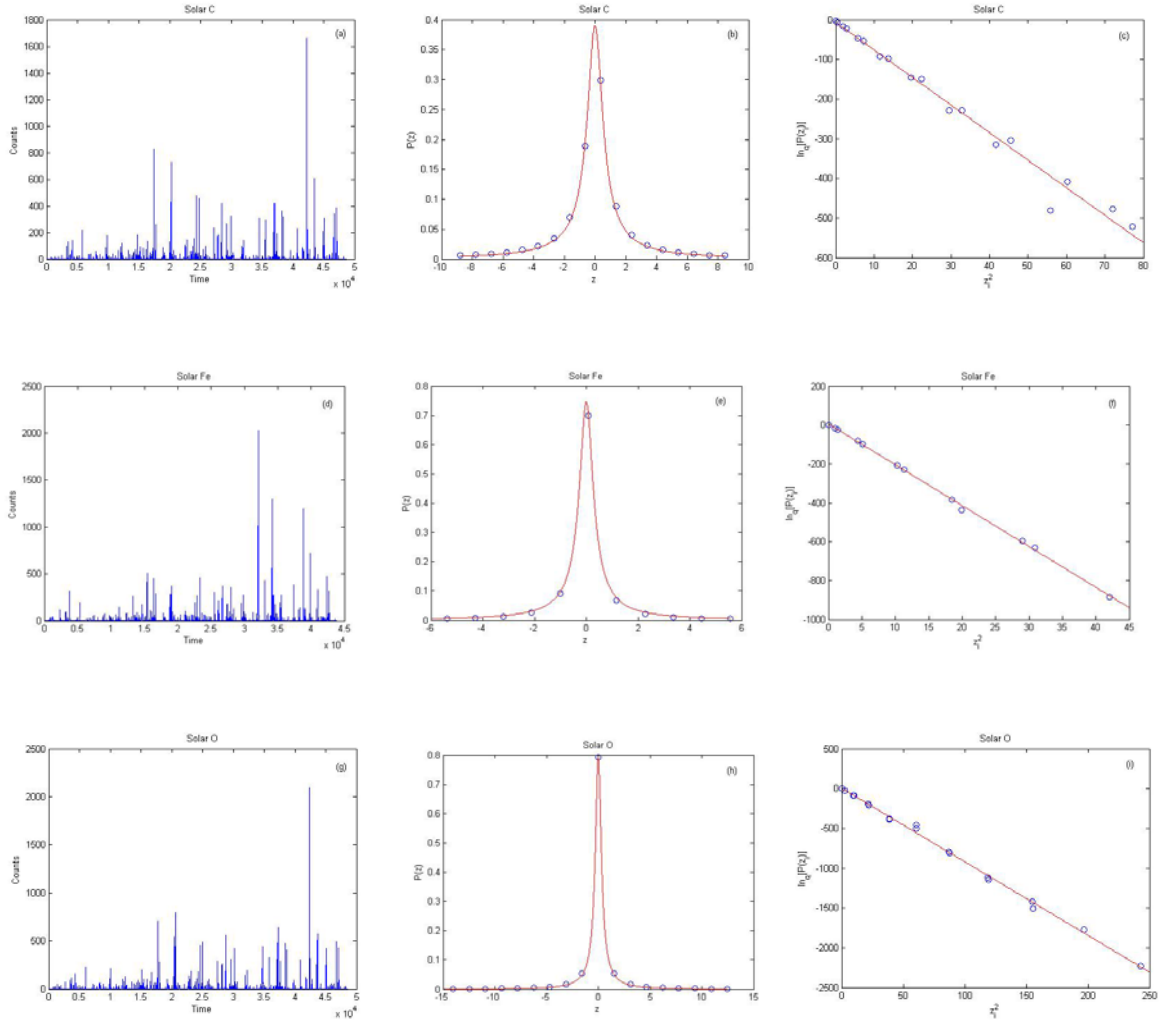

*Figure 7: (a)* Time series of Solar Carbon *(b)* PDF $P(z_i)$ vs. $z_i$ q Guassian function that fits $P(z_i)$ for the Solar Carbon *(c)* Linear Correlation between $\ln_q P(z_i)$ and $(z_i)^2$ where $q = 2.27 \pm 0.05$ for the *Solar Carbon* *(d)* Time series of Solar Ferrum *(e)* PDF $P(z_i)$ vs. $z_i$ q Guassian function that fits $P(z_i)$ for the Solar Ferrum *(f)* Linear Correlation between $\ln_q P(z_i)$ and $(z_i)^2$ where $q = 2.26 \pm 0.03$ for the *Solar Ferrum* *(g)* Time series of Solar Oxygen *(h)* PDF $P(z_i)$ vs. $z_i$ q Guassian function that fits $P(z_i)$ for the Solar Oxygen *(i)* Linear Correlation between $\ln_q P(z_i)$ and $(z_i)^2$ where $q = 2.14 \pm 0.06$ for the *Solar Oxygen*.

### 4.4.5 q-statistics of cosmic star brightness

In the following we study the q-statistics for cosmic star brightness. For this we used a set of measurements of the light curve (time variation of the intensity) of the variable white dwarf star PG1159-035 during March 1989. It was recorded by the Whole Earth Telescope (a coordinated group of telescopes distributed around the earth that permits the continuous observation of an astronomical object) and submitted by James Dixson and Don Winget of the Department of Astronomy and the McDonald Observatory of the University of Texas at Austin. The telescope is described in an article in The Astrophysical Journal (361), p. 309-317 (1990), and the measurements on PG1159-035 will be described in an article scheduled for the September 1 issue of the Astrophysical Journal. The observations were made of PG1159-035 and a non-variable comparison star. A polynomial was fit to the light curve of the comparison star, and then this polynomial was used to normalize the PG1159-035 signal to remove changes due to varying extinction (light absorption) and differing telescope properties.

Figure 8 shows the estimation of q-statistics for two cosmic stars (PG-1159-035) and BCR-681). The q-values for the star PG-1159-035 time series was found to be $q_{stat} = 1.64 \pm 0.03$ and for the star BCR-681 the q-value was found to be $q_{stat} = 1.36 \pm 0.04$ correspondingly. For both cases we clearly observe non-Gaussian statistics.

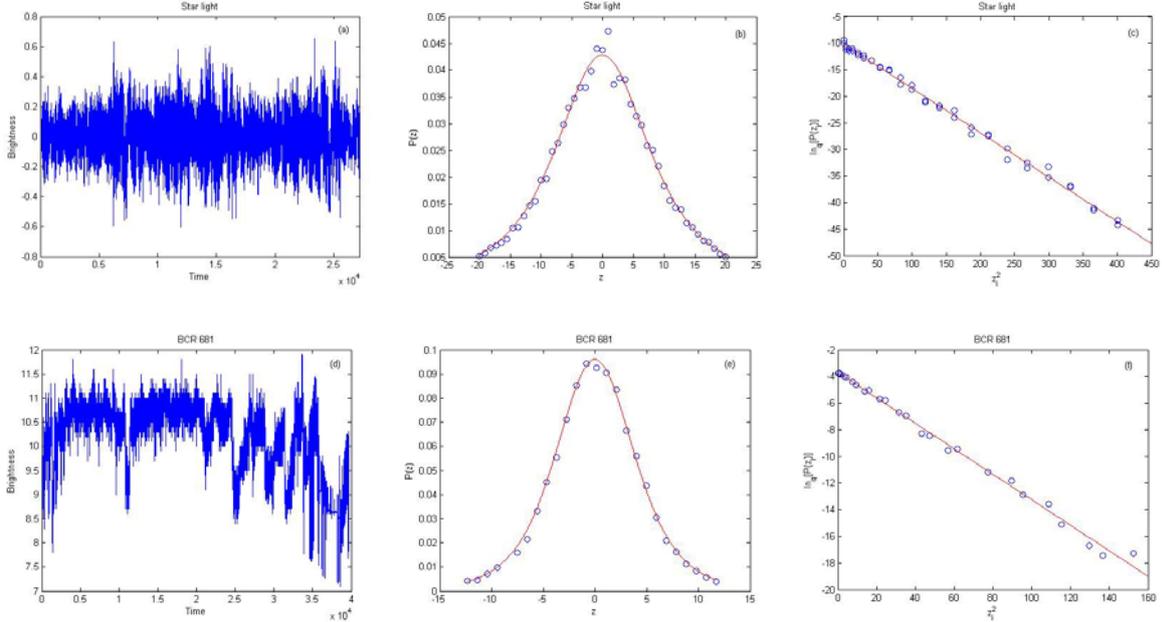

***Figure8:*** *(**a**) Time series of cosmic star PG-1159-035 (**b**) PDF P($z_i$) vs. $z_i$ q Guassian function that fits P($z_i$) for the cosmic star PG-1159-035 (**c**) Linear Correlation between $\ln_q P(z_i)$ and $(z_i)^2$ where q = 1.64 ± 0.03 for the cosmic star PG-1159-035 (**d**) Time series of cosmic star BCR-681 (**e**) PDF P($z_i$) vs. $z_i$ q Guassian function that fits P($z_i$) for the cosmic star BCR-681 (**f**) Linear Correlation between $\ln_q P(z_i)$ and $(z_i)^2$ where q = 1.36 ± 0.04 for the cosmic star BCR-681.*

### 4.4.6 q-statisitcs of cosmic rays

In this sub-section we study the q-statistics for the cosmic ray (carbon) data set. For this we used the data from the Cosmic Ray Isotope Spectrometer (CRIS) on the Advanced Composition Explorer (ACE) spacecraft and especially the carbon element (56-74 Mev) in hourly time period and time zone duration from 2000 – 2011.The cosmic rays data set is presented in Fig.9a, while the q-statistics is presented in Fig.9[b,c]. The estimated $q_{stat}$ value was found to he $q_{stat} = 1.44 \pm 0.05$. This resulted reveals clearly non-Gaussian statistics for the cosmic rays data.

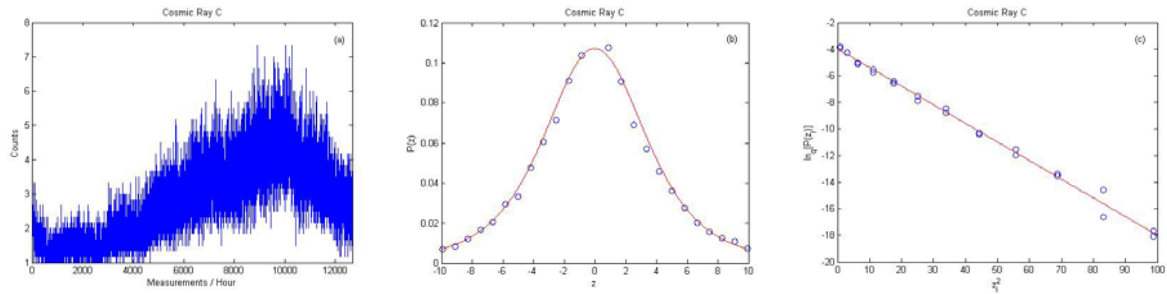

***Figure9:*** *(**a**) Time series of cosmic ray C (**b**) PDF P($z_i$) vs. $z_i$ q Guassian function that fits P($z_i$) for the cosmic ray C data (**c**) Linear Correlation between $\ln_q P(z_i)$ and $(z_i)^2$ where q = 1.44 ± 0.05 for the cosmic ray Carbon.*

## 5. Summary and Discussion

In this study we have presented theoretical and observational indications for the new status of non-equilibrium space plasma theory. Especially, the Tsallis non-extensive $q$-statistics have been verified in every case: the magnetospheric plasma, the solar wind, the solar plasma (convection zone, solar corona) for cosmic stars and cosmic rays. As the Tsallis $q$-statistics reveals long range correlations and strong self-organization process we presented also the highlights of non-equilibrium statistical theory of random fields putting emphasis to the theory of $n$-point correlations, which are responsible for the complex character of the non-equilibrium dynamics. In the following we present further discussion of the space plasma complex processes in relation with the general transformation of modern scientific concepts included in the complexity theory.

For all cases the statistics of experimental space plasma signal was found to obey the non-extensive q-statistics of Tsallis with high q-values ($q_{stat} \simeq 1.50-2.50$). Furthermore, the q-statistics for the magnetospheric system was estimated to obtain q-value ($q_{stat} \simeq 2.00$) for MHD signals (magnetic field bulk plasma flow) and $q_{stat} \simeq 2.50$ for electric field and energetic particles. For solar flares activity and solar energetic particles the q-values were also found to be higher than two ($q_{stat} \simeq 1.90-2.30$). For the solar convention zone (sunspot index) the q-value was found to be $q_{stat} = 1.53$, while for the cosmic rays (carbon experimental data) the q-value was found to be $q_{stat} = 1.44$. The above q-statistics values reveal clearly that space plasma dynamics obeys Tsallis non-extensive statistics in every case, from planetic magnetospheres to solar plasma and solar corona as well as to cosmic stars and cosmic rays.

These results are in wonderful agreement and harmony with the theoretical framework presented in previous sections of this study. In fact, the presence of q-statistics indicates clearly the necessity of fractal generalization of dynamics for the theoretical interpretation of non-equilibrium space plasma dynamics in accordance with many theoretical studies that are cited here. Finally, we believe that the Tsallis q-statistical theory coupled with the non-equilibrium fractal generalization of dynamics indicates a novel road for the space plasmas science. Moreover, this point of research view highlights the space plasmas system as one of the most significant case for the application of new theoretical concepts introduced by the complexity science during the last two or three decades.

## Acknowledgements

We thank the ACE MAG instrument team and the ACE Science Center for providing the ACE data [74]. Moreover, we thank S. Kokubun and T. Mukai for the high resolution Geotail/MGF magnetic field and Geotail/LEP plasma data, respectively.